# Joint Radar and Communication: A Survey

Zhiyong Feng[1,*], Zixi Fang[1,*], Zhiqing Wei[1], Xu Chen[1], Zhi Quan[2], Danna Ji[1]

[1] Key Laboratory of Universal Wireless Communications, Ministry of Education, Beijing University of Posts and

Telecommunications, Beijing 100876, China

[2] College of Electronic and Information Engineering, Shenzhen University, Shenzhen 518060, China

* The corresponding author, email:fengzy@bupt.edu.cn, fangzx@bupt.edu.cn

**Abstract**: Joint radar and communication (JRC) technology has become important for civil and military applications for decades. This paper introduces the concepts, characteristics and advantages of JRC technology, presenting the typical applications that have benefited from JRC technology currently and in the future. This paper explores the state-of-the-art of JRC in the levels of coexistence, cooperation, co-design and collaboration. Compared to previous surveys, this paper reviews the entire trends that drive the development of radar sensing and wireless communication using JRC. Specifically, we explore an open research issue on radar and communication operating with mutual benefits based on collaboration, which represents the fourth stage of JRC evolution. This paper provides useful perspectives for future researches of JRC technology.

**Key words**: joint radar and communication; system integration; survey; collaboration; 5G/B5G

## I. INTRODUCTION

Radar and wireless communication are two of the most common and important applications of modern radio frequency (RF) technologies. Radar is used to detect and identify targets, while communication is utilized to achieve information transmission between devices. Generally, they are independently developed and designed based on their respective functions and frequency bands. Therefore, they do not affect each other [1]. However, with the development of technology, radar and communication are towards joint design. The reasons can be summarized as follows. Firstly, with the exponential increase of wireless devices and amount of data traffic, spectrum is becoming increasingly scarce [2]. Secondly, with the increasing complexity of electromagnetic environments, the performance of single radar is limited. Multiple radars need networking through communication techniques to obtain rapid fusion of large amount of detection data. Particularly, there is a strong demand for integration of radar and communication in some scenarios where space and power are extremely limited, such as vehicle-to-vehicle networks, flying ad-hoc networks and fleets. Thirdly, radar may assist communication to achieve faster neighbor discovery, more accurate channel estimation and beam alignment. As radar and communication are required to satisfy the demands of spectrum efficiency, performance gain and multiple tasks, the development of JRC has gained a lot of attention.

Generally, the principles of radar and communication are different. Radar radiates electromagnetic wave to irradiate the target and receives its echo, so as to obtain the distance, velocity, azimuth, height and other physical state information of the target [3]. One hand, the radar signal is known for transmitter and the target characteristics are desired to be estimated. However, the communication signal of a transmitter is usually stochastic and the transmission channel can be estimated previously based on the data carried in the channel estimation procedure [4]. On the other hand, radar needs as precise channel information as possible to estimate the target, but communication only needs the primary information of channel and some of it may even be omitted. Despite the differences in principle between the two, there still exists possibility of joint design.

With the development of radar and communication technologies, the difference between them has been gradually reduced [5],[6]. Firstly, the signal processing module of radar, traditionally implemented by analog hardware devices, is being replaced by digital devices. The digital communication system has been applied for many decades. Therefore, both radar and communication have strong similarity and realizability in hardware structure and system components [7]. Secondly, phased array is one of the key technique during the development of radar system. The antenna array of a phased array radar consists of a number of radiation and receiving units (called elements) which are arranged in a regular plane to form an antenna array. Multiple input and multiple output (MIMO) is also a multiple antenna system that has been used for communication in recent years. Hence, radar and communication are increasingly similar in antenna structure. From that point of view, the development of MIMO technology also brings opportunities for the integration of radar and communication [8]. Thirdly, the bandwidth of communication is getting wider and even close to the bandwidth of radar. This is due to the utilization of mid-high frequency which brings more accessible spectrum resources for wireless communication recently. Benefited from the high similarity of the two system in hardware component architecture, antenna structure and working bandwidth, it is positive to see that the joint design of these two systems is feasible.

The early research of joint radar sensing and wireless communication can be traced back to the Space Shuttle program of National Aeronautics and Space Administration (NASA) in the 1970s [9]. Recent systematic developments in the field of JRC are motivated by the Shared Spectrum Access for Radar and Communications (SSPARC) project of Defense Advanced Research Projects Agency (DARPA) in 2013 [10], which lays the foundation of jointly designed methods for radar and communication. The research process of JRC includes cognitive radar radio [11],[12] schemes, time/frequency/beam sharing schemes [13], integrated signal processing technologies [14], joint waveform design methods [16]-[18], software defined networks (SDN) [19],[20], non-orthogonal multiple access (NOMA) [21], reinforcement learning [22], etc. The underlying operation method of a joint system should satisfy the requirements for target detection and information transmission simultaneously, and should save resources as much as possible. Recently, there is a widely recognized paradigm to categorize the levels of JRC technology, namely, coexistence, cooperation and co-design [23]. In this paper, we briefly review and summarize the state-of-the-art development of these three categories. Then the characteristics and advantages of JRC systems are investigated. Finally, the applications currently being studied or those benefited from JRC technology in the future are introduced. Particularly, we extend the categorizations in [7] to a fourth mode of operation named collaboration. Then we propose a new architecture for JRC that enables multiple radars cooperative detection.

The remainder of this paper is organized as follows. Section II introduces the basic concepts, characteristics and advantages of JRC. In Section III, we present the applications that can benefit from JRC technology currently and then discuss the future scenarios that may utilize JRC as a key technology. Section IV reviews the state-of-the-art for JRC system and the existing methods of JRC, namely, coexistence, cooperation, co-design and collaboration. Open research problems on the JRC system and potential solutions will be discussed in Section V. Section VI concludes this paper.

## II. CONCEPTS, CHARACTERISTICS AND ADVANTAGES

This section introduces the basic concepts, characteristics and advantages of JRC technology.

### 2.1 Concepts of JRC

JRC implies that the two subsystems can share the hardware platform to achieve different desirable functions at the same time, without causing negative influence on each other. From the process of technological, JRC technique can be divided into two levels: signal and networking.

The previous concepts of JRC mainly focused on the level of signal. The representative technologies include time sharing, spectrum sharing, beam sharing and waveform design. Time sharing is the simplest approach in all solutions for radar-communication integration. In this approach, the communication signals can be transmitted using the time-slot of a certain detection pulse. Furthermore, we can simply reuse the antenna and transceiver by adding a strobe switch. When detection is needed, system switches to radar modem, while communication functions are disabled, and vice versa. Because radar and communication work alternately in the time domain, the mutual interference between them can be avoided and the hardware can be co-used. In the approaches of spectrum sharing, radar and communication signals are transmitted at different frequency domains. For example, by using the principle of orthogonal LFM signal, the multi-carrier LFM signal can be constructed and the carriers used for communication and detection will be arranged alternately. Moreover, it can achieve hardware reuse by utilizing FDA-MIMO technology. Due to the limitation of time-slot and spectrum resources carrying the wireless traffics, the main concern of researchers is the compatibility of radar and communication functions and the objective is to assist improving the efficiency. It is automatically conceptualized as coexistence in this stage and it indicates that there is without any sharing of information within the two subsystems. However, the content of signal extends far beyond time sharing and spectrum sharing, and has evolved into a wide concept in recent years. As a result, researchers have started to investigate not just the methods of coexistence, but also fundamental schemes of the mutual operation and/or design. The typical definition of JRC in modern integrated systems is called cooperation and co-design, which was proposed in [22]. Here, cooperation indicates that some knowledge of radar and communication subsystems are shared by each other and they are not independent anymore. The typical technique is beam sharing. Co-design implies that the previous radar and communication systems will be re-designed. Some researchers suggest that co-design should include waveform design, signal design and hardware design [7].

Networking can be regarded as the future concept of JRC. Indeed, with the explosive growth of wireless devices, JRC should not only be constrained in single user or two-user situation, but also in a number of devices, in which each device shares its information with others through wireless links and the information could be aggregated [24]. Besides, these devices forming local networks must be coordinated and cooperative such as in vehicle to everything (V2X) networks. Hence, we regard the collaboration as an appropriate feature for the future JRC system and the details will be discussed in Section IV-D.

## 2.2 Characteristics of JRC

JRC utilizes technical methods to achieve hardware device reuse and software function redefinition, and further achieves the cost minimization of dual-function. In detail, the JRC system has three important characteristics as follows.

- *Architecture unification and simplification*. The JRC design ensures the integrity and unity of the overall architecture, and has flexibility and simplicity for operation.
- *Functional reconfiguration and scalability*. JRC design has the capability to support software-defined functional reconfiguration and resource integration to ensure complementarity of radar and communication functions.
- *Efficiency enhancement and cost reduction*. Especially in miniaturized carriers, a JRC system has the characteristics of leveraging high efficiency and low energy, in which each carrier can enhance comprehensive abilities and reduce costs substantially.

## 2.3 Advantages of JRC

A JRC system is expected to have significant advantages in terms of equipment cost and size, system performance, spectrum usage. The system utilization should be improved prominently and the system complexity be reduced effectively [25],[26]. We can understand it from three aspects. Firstly，the characteristics of radar are sufficiently shared by the communication subsystem, such as high power, low side-lobe, strong directivity and anti-interference. These advantages can improve communication quality, increase the information security and extend the range of communication. Secondly, the stable transmission has generally provided many substantial advantages for radar such as low delay, high rate, simple implementation and fast response. Thirdly, it is possible to improve the automation level of dual-function systems, and hence, the process of manual handling can be simplified.

One example revealing the advantages of JRC in commercial applications is the next generation automobiles. The integrated approaches would allow vehicles to simultaneously harvest the benefits of autonomously sensing the driving environment and cooperatively exchanging information such as velocity, braking, and entertainment contents with other vehicles instead of relying on human decision [27]. Another example is the reconnaissance mission. One single integrated system means that higher reliability and faster information gathering because there is no manual intervention [13]. Obviously, JRC systems will provide advantages in modern military applications through decreasing the volume, weight, and power consumption of facilities and improving information fusion in an enhanced warfare network.

## III. JOINT RADAR AND COMMUNICATION APPLICATIONS

Before discussing the existing research work on JRC

systems, we first list the various applications that have benefited from the JRC technology currently or could get development from it in the future. Generally speaking, these applications are divided into two parts, namely, military and commercial applications. In this section, we take three military scenarios as examples to illustrate the application process of JRC technologies, which are naval vessels electromagnetic system, avionics system and group battling system. Then, we take three commercial scenarios as examples to discuss the commercial JRC applications in detail, which are intelligent transport system (ITS), drone surveillance network, and air traffic management. The devices in these scenarios are usually limited by space and power.

### 3.1 Military Users

In the future integrated battlefield of sea, land, air and space, military platforms such as tanks, warplanes, and warships will be loaded with diverse electronic devices to enhance their interoperability and probability of survival [28].

#### 3.1.1 Naval Vessels Electromagnetic System

Modern navy has raised higher requirements for the function, performance and interoperability of shipborne combat systems. However, the increasing volume of electronic equipment such as radar and communication subsystems on warships have made antennas highly congested on decks and bridges. Furthermore, such a crowded environment means that each RF system usually has its own antenna aperture, inevitably leads to electromagnetic interference or compatibility problems. Therefore, an integration solution is urgently needed.

The Office of Naval Research (ONR) initiated an Advanced Multifunction RF System (AMRFS) program to address and resolve the RF functions integration challenges for US Navy in 1996 [29]. The goals of the program are to reduce the number of RF system apertures dramatically while increasing effective functionality and bandwidth, and further reduce the aggregate radar cross section (RCS) to enhance the combat capability [30]. The realization of AMRFS experimental platform proves that multi-tasking of RF electronic resources is a feasible approach. This experimental platform can complete detection, navigation, communication, electronic warfare and other functions on a common RF module. According to the task requirements, the system forms a single beam or multiple beams with specific spatial direction to switch the beams in real time [31]. In 2009, ONR initiated the Integrated Topside (InTop) program to develop a range of scalable multifunctional aperture supporting electronic subsystems. InTop program, the upgraded version of AMRFS, can achieve radar and communication functions on an integrated architecture where the RF functions share apertures, electronics, displays and operators are all controlled on a priority bases by a common resource allocation manager software [32]. Given that rigorous task requirements, ONR further puts forward the Electromagnetic Maneuver Warfare Command and Control (EMC2) program aiming to improve the integration of electronic equipment and command/control systems. It provides a prospect for the next generation integrated multifunctional shipboard RF system. A related work is the cooperative robotic surveillance networks, in which the vessels complete command and control instruction via communication subsystems to support schedule missions for controllers [33].

#### 3.1.2 Avionics System

In warplane avionics system, the JRC technology has been used for decades. Joint design is the core of avionics system, which can reduce the number of airborne equipment so as to lower RCS of warplane, and reduce the work load of pilots. Therefore, it can reduce the maintenance cost, and further improve the combat effectiveness. A typical joint avionics system is PAVE PACE of U.S. Air Force and has been used in F-35 fighter, which owns the ability of jointly modular processing. That makes it possible to achieve both radar and communication functions by sharing RF aperture, such as the super-heterodyne receiver. In this case, the integrated sensor system (ISS) could also benefit from general module, resource-configurable and opening testability environments in the next generation avionics system.

#### 3.1.3 Unmanned Group Battle System

With the increasingly diverse war demands, an unmanned group battle system provides an effective method for cooperative reconnaissance, group attack, electromagnetic countermeasure, tactical feint and other modern combat modes. However, the unmanned group battle system needs real-time environment sensing and rapid information interaction to guarantee the success of missions. The JRC systems will be able to play an important role in the future unmanned group battle system.

The potential unmanned group battle scenarios applying the JRC technology include unmanned aerial vehicle (UAV) groups, tactical unmanned vehicle (TUGV) groups and missile Ad Hoc networks. This is specifically an interest of the DARPA Gremlins programs and it utilizes the airborne radar to perform the detection and share detection information within the UAV groups simultaneously. After that, the OFFensive Swarm-Enabled Tactics (OFFSET) program proposed by DARPA has been launched aiming to combine UAV with TUGV to improve the battle efficiency in urban environment. The U.S. military has also launched some research programs related to missile Ad Hoc networks whose information sharing scheme is jointly designed, such as Gray wolf program proposed by Air Force Research Laboratory (AFRL) and LOCUST program proposed by ONR.

Based on the program of SSPARC, an approach of comparing various cooperative schemes for spectrum access has been proposed, and it can be used for minimizing the impact of sharing spectrum to support the demands of inter-group communications [34]. The forward channel signal-to-noise maximization while simultaneously minimizing co-channel interference has also been considered for a joint consideration and operation of SSPARC [35].

## 3.2 Commercial Users

The commercial scenario is another potential application scenario that regards JRC as a promotion of advanced technology, and it has been brought into focus recently. The typical scenarios include the intelligent transportation system, the drone surveillance network, and the air traffic management system.

### 3.2.1 Intelligent Transport System

JRC is a good candidate technology for Intelligent Transportation System (ITS). The key motivation is its ability to enhance the safety of transportation, which means the safety of drivers and pedestrians can be further guaranteed through enhancing the reliability of communication and utilizing radar signal or expanding the detection range of radar using communication packets. For example, the information of transportation can be exchanged rapidly, accurately and instantly, especially in those complex traffic situations, such as intersections. On the other hand, future vehicles will act as an important enabler for communication connectivity from a single node to a large number of nodes on road according to the vision of 5G/B5G V2X.

In addition to the consideration of improving security, the demand of data rates in vehicle terminal is also increasing. This is mainly represented in the huge original data of fully automated driving, high-precision 3D navigation maps, interactive entertainment information [26]. Therefore, the idea of joint automotive radar and communication (JARC) was proposed [36] and the vehicles with millimeter wave radars were exploited to realize the high data rates in ITS [37]. Solutions based on a communication standard such as 802.11 are original for implementing simultaneous JARC systems at the 60 GHz unlicensed band [26]. With this assumption, an approach that determines the mean-normalized channel energy by frequency domain channel estimate was presented in [38] to enable closest target range estimation. Moreover, a processing algorithm of radar receiver for range and velocity estimation was developed in [39]. Related techniques such as waveform design [40] and beamforming design [41] based on 802.11ad for JARC systems were also studied. In addition, it is worth mentioning that the most prevalent communication standard of vehicle network is short-range communications (DSRC) which was proposed by the United States Congress in 1998. This standard is based on an evolution of IEEE 802.11p and it operates at 5.85-5.925 GHz. Obviously, this spectrum cannot support a large amount of data, which is on the magnitude of Gbps in ITS [42]. However, it can still achieve both radar and communication functions if we employ orthogonal signals based on IEEE 802.11p standard [43]. Other researches of JARC based on orthogonal frequency division multiplexing (OFDM) technology include waveform design scheme combining differential phase shift keying (DPSK) communication symbols with phase modulated continuous waves (PMCW) radar sequences [44], subcarrier allocation techniques for multiple-user access, and direction of arrival estimation based on vehicle radars [45],[46].

The JARC hardware solutions have also drawn much interest recently. For example, 24 GHz band using the emerging substrate integrated waveguide (SIW) technology was applied in an integrated radio and radar system through a single transceiver platform towards millimeter wave applications of ITS [47],[48]. Multi-functional transceivers for future intelligent transportation platforms were developed to sustain the highly mobile high-speed communication and high-resolution radar sensing [49]-[52]. Additionally, a joint mmWave MIMO radar and communication solution with low resolution analog-to-digital converters per RF chain was proposed for supporting the requirements of next-generation vehicles [53].

In order to realize the JARC systems, researchers implemented a lot of measurements in laboratory and actual road environments. A JARC test platform with laboratory environments was deployed in [54], it utilizes the 802.11 OFDM communication waveform, as found in IEEE 802.11a/g/p, to perform forward collision warning (FCW). This scheme achieves 1 m accuracy for a single target with bandwidth of 10 MHz. The MIMO-OFDM integrated JARC network in a vacant parking was established, and the performance of data transmission, the accuracy of detection with interference was verified, respectively [37]. In test environment of field, the feasibility of long-term evolution (LTE)-based passive radar for vehicle detection and tracking was tested [55]. Compared with field test, an alternative software simulation approach has the advantages of low cost, time saving, and repeatability. An useful tool, called Virtual Drive, which consists of a traffic model and the calculation of the multipath propagation with a 3D ray-tracing tool was proposed [56]. It provided a feasible way to optimize the antenna configurations for sensing and communication at the same time. Overall, the JARC system will be positive to support the automated driving and ITS equipment.

### 3.2.2 Drone Surveillance Network

Drone surveillance radar network is another potential commercial application of JRC technologies. Recently, drones have been applied widely in many fields, such as geographic exploration, disaster relief, logistics transportation, emergency communication, which brings tremendous convenience to electronic consumers [57]. However, the development of drones also imposes great threats to public safety and security, especially those drones with hostile purposes [58]. Thus, the surveillance and control of drones have gained a lot of focus and radar is a promising approach for detection and tracking of drones. In particular, those drone surveillance radars that have the advantages of mobility, miniaturization and rapid deployment are very effective against low altitude, low flying speed and small-size drones. Moreover, the number of drone surveillance radars is bound to increase greatly in order to adapt to the construction of smart cities. To get accustomed to this situation, the Single European Sky ATM Research Joint Undertaking (SESAR-JU) has developed the U-Space framework to ensure the safety of drone traffic management, and the drone surveillance radars will be widely deployed in various places [59]. Other institutions have also introduced radar-based programs for drone traffic management, such as NASA, METI, and ICAO [60].

Currently, the products of drone surveillance radars not only have the advantages of operating all the time with high precision and resolution, but also can realize the information transmission and interaction with other radars simultaneously through establishing communication links [61],[62]. This is also a trend for the future drone traffic management system and provides prospects for JRC technologies.

### 3.2.3 Air Traffic Management System

Due to the rapid growth of commercial aircrafts, the air traffic control services for navigation and connectivity are undergoing an extensive modernization process [63], and introduce the integration techniques of sensing and communication. Automatic dependent surveillance broadcast (ADS-B) is a candidate JRC surveillance technology, in which an aircraft broadcasts its various flight parameters (e.g. latitude, longitude and level) periodically, enabling itself to be tracked by other aircrafts in the vicinity [65]. Furthermore, ADS-B can provide a radar-like display of aircraft for ground controllers instead of multi-radar deployment and thus allow the control of aircraft in radar-blind area such as ocean. Thus, the accuracy of related data is no longer susceptible to the position of aircraft or length of time between radar sweeps.

To simplify deployment and reduce costs, some researches have utilized L-band digital aeronautical communication system type 1 (LDACS1) communication signals as signals of opportunity for the passive MIMO radar implementation [66]. Others have focused on the multi-function waveforms, which were experimented on an Air Force Research Laboratory (AFRL) software-defined radar testbed [68]. Implementation of an OFDM architecture platform was targeted at low weight in the airborne radar sensor network environment [69]. Moreover, other cases of air traffic management such as traffic collision avoidance (TCAS) and air traffic control (ATC) also benefit from JRC technologies.

# IV. STATE OF THE ART OF JRC

In this section, we discuss the state-of-the-art of JRC techniques according to three categories: coexistence, cooperation, and co-design proposed in [22]. We further discuss a fourth category of JRC called collaboration. For each category, we will review the detailed research methods and discoveries in the past few years. This facilitates to form a comprehensive exploration outline of the JRC technology. For clarity, the comparison of different categories are shown in Table I.

**Table I.** *The comparison of different categories.*

| Stage | Mode | Implication |
|---|---|---|
| Co-existence | Uncoordinated | Radar and communication are physical isolation and treat each another as the interferer. |
| Cooperation | Coordinated or Uncoordinated | Two independent systems can operate at the same level, and exploit the joint knowledge to promote the performance of both systems. |
| Co-design | Coordinated | A complete joint unit which including modern integrated technologies such as signal, waveform, coding, etc. |
| Collaboration | Completely coordinated | Multiple complete joint units networking and work together to accomplish common tasks, radar and communication will be mutual promotion. |

## 4.1 Coexistence

As mentioned above, radar and communication transceivers treat each another as the interferer in coexistence. Therefore, it is necessary to analyze the effect that each system has on each other [70].

In order to analyze the impact of radar systems on communication, HAN *et al.* in [71] proposed a numerical calculation with conditional probability to obtain the bit error rate (BER) of 2.4 GHz LTE system under the interference from radar. It also suggests that some measures for interference mitigation must be used such as distance separation or frequency separation. The authors in [72] provided an analysis model to study the error rate performance of uncoded real-valued modulation schemes under three cases: treating interference as Gaussian noise, using interference cancellation scheme and a worst interference environment. In [73] and [74], two system level analyses were performed to analyze the interference of radar systems to LTE systems in 3.5 GHz. Both of them indicate that although the interfering signal of pulsed radar decreases signal-to-interference-plus-noise-ratio (SINR) of communication, the performance of LTE eNodeB will be seriously influenced only when the distance of interfering transmitters using the overlapping frequencies is very close. On the other hand, the impact of communication systems on radar performance was analyzed in [75]-[78]. According to the theoretical derivation [72] and the measurements, it is confirmed that the interference imposed to radars by communication systems will degrade the performance of radars. Consequently, the significant isolation distance between LTE base stations and radar stations was calculated. Furthermore, a mathematical model for the analysis of radar performance under cellular interference was proposed in [75]. The results prove that the sum of interference signals from a cellular system has a log-normal distribution and provide a useful interference distribution. Others focused on the comprehensive performance of integrated system. The compound rate was proposed to optimize the degrees of freedom of radar waveform and the encoding matrix of communication symbols under constraints regarding to SINR and power [79]. An achievable performance bound of radar and communication of coexistence system was studied and the appropriate integrated system which depends on the requirements of users can be designed by utilizing the derived inner bound [80].

These literatures reveal the mutual interference between radar and communication. However, the interference cancellation schemes to reduce the mutual interference are necessary. A joint precoder-decoder design was proposed to maximize the SINR for coexistence JRC systems [81] and the Interference Alignment (IA) constraints was proved to be effective for facilitating interference cancellation. Researchers addressed the problem of interference cancellation as a joint waveform estimation/data demodulation problem, especially in uncoordinated scenarios of radar and communication coexistence [83]. For coordinated scenarios, time sharing methods is a direct and simple implement solution avoiding mutual interference and reducing system complexity. For example, using an electronic switch and an timeslots assignment for a single antenna can achieve channel reuse [84],[85]. However, this implementation only works in pulse radar mode, and the slots for communication will interrupt the radar detection that could lead to targets missing. Moreover, the system efficiency is low due to the waste of time in waiting for a target echo in the timeline. Hence, more

The final version of this paper is published in China Communications    7

high-efficient methods are desirable.

When radars adopt the MIMO technology, it becomes more resilient for waveform shaping and beam alignment. Consequently, the arbitrary interference generated by wireless systems from any directions can be minimized [86]. Furthermore, if the joint schemes in MIMO radar and communication systems are designed well, the coexistence can achieve maximized SINR at the radar receiver and the required communication rate simultaneously [87]. As a trend towards future flexible and intelligent system such as 5G structure, combining MIMO radar with MIMO communication to develop a joint system that has the ability of accurate sensing and high-speed transmission has also been investigated. The authors in [88]-[90] proposed the concepts of MIMO matrix completion (MIMO-MC) employing sparse sampling to reduce the amount of data in radar fusion center under certain communication power and rate constraints. Other researches focused on the waveform control using the degrees of freedom of antenna array. For example, a balance between the SINR of radars and the rate of communication was optimized through a reduced-complexity iterative algorithm in the radar transmit array [91],[92]. Based on this algorithm, [93] provided the closed-form and fast convergence transmit approach for the coexistence of MIMO radar and MIMO communication system. Similarly, researchers in [94] first split the antennas into two groups, one for radar and the other for communication. That makes the waveform control algorithms reliable. Based on the partitioning technique proposed in [90], a JRC capable monostatic coherent MIMO radar system was developed. It reveals several advantages such as fully coherent signal processing and coherent transmit beamforming that minimizes the intercept probability of beam patterns [95]. Furthermore, some researches attempted to mitigate the clutter within a coexistence JRC system using precoding technologies [96],[97].

Employing the existing cellular networks has been considered as another method to solve the coexistence of two separated subsystems [98]-[100]. These approaches regard the base station (BS) as a joint transceiver hardware platform, which can guarantee either the receive SINR level of downlink users or the interference-to-noise ratio (INR) level to radar. Especially, in [94], the probability of detection ($P_D$) of radar was maximized under the constraints of quality of service (QoS) of users and transmit power at the cellular network. The interference between 3.5 GHz 5G network and the radar operating in adjacent band was tested to support the feasibility of coexistence [101]. In addition, some work focused on the hardware methods of coexistence. With the attention of white space radio spectrum, some researchers have studied the problem of designing the hardware platform in a commensal radar system using the emissions of the IEEE 802.22 standard [102],[103].

## 4.2 Cooperation

Cooperation means that two independent systems can operate at the same level, not only referring to the sharing hardware platform but also the objects whose locations are geographically fragmented from each other. Moreover, the two systems no longer regard each other as interferer, but exploit the joint knowledge to promote the performance of both systems [22]. From this perspective, cooperation is actually the first step toward the jointly designed system.

However, the cooperation performance of a dual-function framework will be subject to the characteristic of both subsystems. For example, due to the high power transmitted by radar to detect the far targets, the communication receivers could be damaged when receiving these high power waves. Conversely, the smaller radar echo will enter into the receivers along with the desired communication signal, which could decrease the signal-to-interference-ratio (SIR) of radar and increase the complexity of demodulation and interference cancellation. To solve these problems, numerous methods had been proposed to facilitate the analysis of performance of the dual-function cooperation framework. The performance bounds of cooperative radar and communication systems operation was analyzed by a Bayesian framework [104] or a structured covariance-based water filling solution [105]. Researchers also addressed the performance tradeoffs to evaluate the constraint relation in cooperation systems. One example is how to maximize the probability of detection, while satisfying the information rate requirement in the facility consisting of a passive radar and a communication receiver [107],[108]. Similar to the facility-based methods, an optimal power allocation strategy for leveraging the probability of false alarm and the communication rate, which is represented by bits per symbol, was studied in a joint bistatic radar and communication system [109]. Furthermore, a low complexity linear minimum mean square error optimal pilot symbol aided modulation scheme was proposed to enable performance analysis in the same scenario but with single-input single-output (SISO) bistatic radar [110]. Besides, the target localization Cramer-Rao bound and mutual information in a MIMO radar and a distributed MIMO communication system was studied in [111], and the results show that there is a performance gain due to the cooperation in a dual-function system. Other performance analysis of cooperation considering the clutter has also been studied [112].

On the other hand, some researchers attempted to realize the primary detection function via mainlobe while ensuring the transmission of digital communication signals via sidelobe in a dual-function system by using beam control [113],[114] or waveform diversity [115] but still suffered from reduced SNR, cross interference and low throughout. Therefore, the frequency diverse array (FDA) was employed in JRC systems [116]-[118]. The FDA applies an additional linear frequency shift to the elements of array and provides greater degrees of freedom for range and angle control, permitting simultaneous multi-mission operation [119]. Based on this, [120] combined butler matrix with FDA using sidelobe manipulating for dual-function joint systems and the proposed method was proved to be effective for dealing with target detection and communication transmission simultaneously. Similarly, some researchers proposed an aperture sharing approach where the transmit or receive arrays are partitioned into several sub-arrays, and each sub-array serves one function or task [121],[122].

Another way to design the cooperation methods is to use modulation techniques. Some researchers proposed to use sidelobe amplitude shift keying (ASK) for the JRC systems, however these methods are not investigated so far under fading channels [123],[124]. Obviously, the fading environment actually exists in wireless channels due to multipath propagation. Based on this fact, [125] supplemented the fading channels into sidelobe ASK modulation for joint design. Some researchers extended the methods to phase modulation. For example, a novel phase modulation based dual-function system associated with beamforming weight vectors was developed to decode the embedded binary sequence in communication receiver [126]. Combining binary phase shift keying (BPSK) with quaternary phase shift keying (QPSK) to modulate the code bits is also an effective method for improving the range resolution and reducing the range sidelobe levels [127]. Moreover, embedding quadrature amplitude modulation (QAM) based communication information in the radar waveforms was proposed to support the communication transmissions of multiple receivers located in the sidelobe region [128]. Finally, there are examples of cooperation systems employing M-ary position phase shift keying (MPPSK) to maximize the relative entropy of the target, meanwhile to minimize the mutual information of the received signals in joint MIMO transceiver [129]. In addition, some researchers focused on the quasi-orthogonal [130] or orthogonal [131] waveform design for operating the integrated systems. Beyond that, there are some other methods for achieving cooperation, such as power control [132],[133], subscriber techniques [134] and resource allocation [135]. Although these methods are not the real co-design, they still achieve the dual-function in a joint platform.

**4.3 Co-design**

According to the definition given by [22], the co-design methods describe a complete joint system, in which the new system will be projected from ground up, including modern integrated technologies such as signal, waveform, coding, etc.

For co-design, early researches addressed the problems posed by waveform design utilizing linear frequency modulated (LFM) signal and spread spectrum (SS) signal. An investigation of radar and communication utilizing SS waveform is in [136]. Direct sequence spread spectrum technique (DSSS) is an attractive technique which is highly secure and robust so that it can ensure orthogonality between radar and communication [137]. In addition, DS has also been combined with ultra wide band (UWB) radars [138], making it possible to avoid jamming mutually through utilizing different pseudo-noise code to spread the spectrum of radar and communication. Given that a no-jamming requirement or application of co-design system in complex electromagnetic environment, chirped spread spectrum (CSS) was proposed by [139]. A novel compatible waveform produced by modulating LFM signal with minimum shift keying (MSK) for joint signal design was proposed in [140]. Based on linear frequency modulation and continuous phase modulation (LFM-CPM), some researches modified the mapping codebook of communication symbols to avoid the performance loss brought by the attenuation of the power amplifier (PA) and the interference from other devices out of band [141]. However, the transmission rate of integrated waveform using LFM or SS technologies is low, and the parameter adjustment is not flexible. Besides, it is difficult to ensure orthogonality of the integrated signal.

Utilizing the characteristics of high SNR radar signal, the authors in [142] provided a novel RF steganography scheme to conceal digital communication in linear chirp radar signals. Although the data rate of communication for this approach is not high, the communication could allow for some administrative functions, low probability of intercept data transmission, navigation function, etc. [143]. In order to obtain adequate synchronization accuracy, a variable symbol duration design was employed to enhance the security of the hidden communication signal by eliminating the cyclostationary features [144]. However, more effective coding and modulation schemes are needed to improve the communication throughput while achieving both low bit error ratio and high radar performance [145].

Another way to address the waveform design in previous work relied on coding approaches. The theory of the coding in relation to the development of adaptive radar and to the construction of spreading sequences in communication systems was discussed in [146], and it established the foundation for the representation of radar environment in terms of operators on the space of waveform. A Doppler resilient sequence of Golay complementary waveform was investigated which the pulse train ambiguity function was proved to be free with range sidelobe at moderate Doppler shifts in [147]. The results provided enormous benefit for instantaneous radar polarimetry. Some researches considered the joint transmission and channel state estimation problem as a channel coding problem with input distribution constrained by an average estimation cost constraint [148]. However, this equivalent problem adopts the assumption that the transmitter is oblivious to the channel state information. Following the principle of spread spectrum techniques, some researchers intended to facilitate the separation of multiple uses for joint system using complete complementary code family [149]. Finally, encoding the data of radar and communication onto a parameter of a particular random distribution for embedding the data in one single waveform was investigated in OFDM pulses [150].

The JRC waveform using OFDM waveform has better characteristics of low sidelobe, high Doppler tolerance and information transmission capacity. With the wide application of OFDM waveform in modern communication, OFDM based feasible options have been attractive as a solution to adapt the development of JRC. The authors in [151] first proposed the design of an OFDM system that simultaneously performs radar and communication operations in 24 GHz band. Based on measurements, they further verified that a very high dynamic range could be achieved for radar imaging with the OFDM-based approach [152]. This work was extended to cope in a multipath-multiuser environment in [153]. However, the interference of multipath-multiuser environment could severely degrade the dynamic range of radar. Therefore, aiming at improving the SNR of radar under OFDM options, some interference cancellation methods were promoted in [154],[155]. Specially, an interference analysis based on stochastic geometry tools was explored for evaluating the extent of interference [157]. Furthermore, a processing algorithm that allows for estimating the velocity of multiple reflecting objects with standard OFDM communication signals was discussed in [158]. Other methods of OFDM options include waveform design for co-designed systems. The authors in [159] used the target detection performance and the channel capacity as the criterion for OFDM waveform design. Mutual information based OFDM waveform optimization was studied to control the interference caused from cellular systems [160]. Similarly, the conditional mutual information and data information rate based on the integrated OFDM waveform was designed to provide acceptable target classification performance and communication rate [161], while others proposed low probability of intercept (LPI) as the performance metric of optimal waveform design for communication interference control [162], radar noise jamming power allocation [163] and total radiated power minimization [164]. In addition, a phase-coded OFDM integrated waveform based on cyclic shifts of M-sequence was proposed, where corresponding time shift was controlled by the communication data [165].

However, the peak-to-average power ratio (PAPR) of OFDM signal is so high that the distortion of transmitted waveform is unavoidable, which means that the detection range of radar will be reduced seriously [166]. On the other hand, the OFDM waveform is continuous for communication, but not for pulse radar. If OFDM is applied to continuous wave radar, the transmitting and receiving antennas must possess good isolation to reduce self-interference, which is difficult in hardware implementation, extremely for the current miniature antenna.

Otherwise, if pulse signal is adopted, the transmission rate of communication will be reduced seriously. Hence, in order to avoid the problems brought by high PAPR and satisfy the requirements of transmission rate and detection resolution, JRC systems employing MIMO was proposed. Particularly, the new schemes combing MIMO and OFDM techniques attract more and more attentions.

Previously, MIMO was applied to communication systems. Due to the RF similarity between radar and communication transceiver, MIMO technology can also be well applied to radar, which makes MIMO-based JRC an attractive area [167]. For communication, it is capable to improve the transmission rate as well as the SINR through the spatial freedom of transceiver antennas. For radars, the MIMO techniques can extend the size of virtual aperture effectively and achieve multi-beam operating for multiple targets searching, locating and tracking [168]. Ultimately, MIMO radar has the unique advantages of spatial diversity and waveform diversity, which can effectively alleviate fading, improve resolution and suppress interference. Consequently, another way to address the problem is the waveform co-design, which combines MIMO radar and MIMO communication. Note that this solution is different from the category of coexistence, which also uses MIMO technology to sense and adapt one another in an integrated system. In the case of waveform co-design utilizing MIMO technology, it is clear that the joint system must be designed to optimize performance with respect to both functions. Some researchers studied the optimum waveform design problem for target parameter estimation. The general colored noise incorporates the normalized mean square error as a design criterion in addition to the mutual information and minimum mean square error in [169]. Given that an effective beamforming algorithms, MIMO technologies can enable a collocated MIMO radar and a full-duplex (FD) multiuser MIMO cellular system consisting of a FD base station serving multiple downlink and uplink users simultaneously [170]. However, MIMO receivers will suffer from the clutter and electronic interference of multiple paths and channels. The ability of interference suppression determines the performance of target detection and communication transmission in the process of receiving echo [171]. Moreover, in order to suppress the interference effectively and improve the SINR of received signals while realizing high-speed data transmission, recent works have focused on combining OFDM signals with MIMO. JRC utilizing MIMO-OFDM technique has advantages such as multiple paths resistance, high spectrum utilization and flexible system diversity [173]. By optimizing the JRC waveform of MIMO-OFDM, an integrated signal with large time-bandwidth product, low cross-correlation sidelobe and low PAPR can be realized [174]. Some researchers investigated the range and angle estimation methods to match the array aperture and the entire signal bandwidth using MIMO-OFDM waveform design [175]. Other researchers addressed the MIMO-OFDM performance analysis to satisfy the requirements of both radar and communication [176]. The results show that compared with the SISO system, the MIMO-OFDM system can improve the angle resolution of radar and the data rate of communication simultaneously. Finally, there are some literatures on MIMO-OFDM employing data fusion algorithms to cancel mutual interference within JRC systems [178],[179] and the interference cancellation concepts for use in a multiple-user access environment was verified at a scaled-down carrier frequency through universal software radio peripherals in [180]. On the other hand, the future JRC systems with MIMO technology are likely to own large spectrum band. Under this assumption, [181] studied the parameter estimation of an OFDM-based joint wideband SIMO radar and communication system based on coherent multidimensional parameter estimation framework. With the development of signal processing, co-design JRC systems on hardware have also attracted wide attentions. An integrated signal processing hardware platform utilizing dynamic partial reconfiguration of FPGA for integrating radar and communication was proposed in [182]. The transmitting and receiving algorithms subject to the PAPR constraint in radar and communication applications were proposed in [183]. Moreover, a JRC receiver was implemented to verify the feasibility of accomplishing dual-function with a single waveform in [184]. The summary of different waveform techniques are listed in Table II.

**Table II.** *The comparison of different techniques of waveform design.*

| Technologies | | Characteristics |
|---|---|---|
| LFM (e.g., [135,136]) | LFM-MSK | Easily implementing in the existing devices. But need accurate signal separation. Low data load. High costs and complexity. |
| | LFM-CPM | |
| SS (e.g., | DSSS | High security and robustness. Higher anti-jamming capability. But low transmission rate, low flexible of parameter adjustment. Need to guarantee |
| | CSS | |
| [131-134]) | DS-UWB | orthogonality. |
| OFDM (e.g., [146-148,153-159]) | | Low sidelobe, higher Doppler tolerance and information transmission capacity. But unavoidable waveform distortion and incomplete use of radar's power amplifier due to high PAPR. |
| MIMO-OFDM (e.g., [166-172]) | | Good spatial and waveform diversity. Large time-bandwidth product. Low cross-correlation sidelobe and low PAPR. |

**4.4 Collaboration**

The above three approaches to realize JRC mostly focus on the link level integration to realize the fundamental functions of radar and communication. With increasingly tight combination between these two fields, the future JRC technology must be implemented for the aggregated performance of multiple nodes, rather than merely meeting the needs of a single node, which means that the radar and communication should work together to accomplish common tasks [185]. Moreover, collaboration further reveals that the nodes which occupied with joint functions are networking. The representative scenarios include vehicle-to-vehicle networks and flying ad-hoc networks. Therefore, collaboration should be the direction of the next generation of JRC systems. In this section, we discuss the performance gain on collaboration and the scenarios that could benefit from the mutual promotion paradigm respectively.

1) *Performance Gain on Collaboration Systems*

Performance gain on collaboration systems indicates that the two subsystems will gain mutual performance advancement by adopting cooperative methods [186]. In order to analyze the performance of JRC systems, some researchers adopted the Cramer-Rao lower bound to reveal the performance of the estimation for the target position and velocity in a passive radar system using existing global mobile wireless communication system [187]. A modified Cramer-Rao lower bound on the joint estimation was investigated for a passive multistatic radar system with antenna arrays based on universal mobile telecommunication system [188]. Similarity in target estimation performance of passive radar was found by employing several dispersed existing communication transmitters, which could be the base station. The communication signals were modified to be used for target estimation of passive radar [189]. Others applied the JRC performance bounds to interaction gain analysis on collaboration systems. For example, the joint performance bounds in terms of the communication rate and radar estimation information rate were proposed in [190]. To some extent, it provides foundation to evaluate the performance of a JRC system from the perspective of information theory. Besides, researchers extended the joint performance bounds to frequency modulated continuous wave (FMCW) radar achieving local Doppler estimation [191] and global radar estimation rate employing optimal radar waveform [192]. Furthermore, [194],[195] extended the bounds of JRC to the multiple-antenna scenario and derived an outstretched domain from temporal and spectral to spatial. Finally, the joint performance bounds adapting to the real transmission channels environments were studied considering the cluster, which could lead to the degradation of collaboration systems performance [196]. The theoretical analysis provides the feasibility of collaboration in future JRC systems.

2) *Applications based on Mutual Promotion*

The JRC technology allows the two subsystems to treat each other as partner. This means that any information able to improve the other subsystem's performance is shared. For example, communication-aided multi-radar cooperative detection will expand the effective detection range of a single radar. The cooperative detection refers to the case where the locations of radars are different geographically. This case consists of the distributed detection and centralized detection. The typical distributed detection scenario is advanced driving assistant system (ADAs) in Internet of Vehicles where the safety message or raw sense date from Forward Collision Warning (FCW), Traffic Jam Warning (TJM) and Blind Spot Warning (BSW) is quickly exchanged and shared among vehicles, such that the drivers can get sufficient real-time road information to obtain plenty extra time to deal with dangers. Another example for mutual promotion scenario supported by centralized detection is multistatic cloud radar systems. The information collected from different radars are processed at a cloud fusion center through communication backhaul links to improve the target recognizing accuracy [197],[198]. Others focused on the radar information aided wireless communication networking, which uses the target position information detected by radar to accelerate the process of neighbor discovery [199]. Among these mutual promotion scenarios, a novel JRC enabled cooperative detection structure has been proposed and will be presented in details in the next section.

## V. OPEN RESEARCH ISSUES

It is promising that collaboration techniques using the communication/radar principles will be important tools for improving radar/communication performance. In this section, we focus on open research issues in networking of multiple radars driven by JRC technologies. The communication-aided multi-radar cooperative detection has been analyzed. The radar-aided wireless networking, channel estimation and beam alignment have been discussed, respectively. An effective sharing pool model for resource allocation has been described aiming at improving the efficiency of joint systems. Furthermore, multidimensional signal processing and improved information theory has been proposed. We discuss the costs and scalability of JRC in the end. Details of the six parts are as follows.

### 5.1 Communication-aided Multi-radar Cooperative Detection

In cooperative detection of geographically dispersed radars, the connection reliability of multiple nodes is necessary for data fusion. Therefore, it is still an open issue to formulate and solve the channel reliability problem.

Recently, researchers use the successful probability of communication defined by SINR to evaluate the quality of wireless channel. It makes this problem analyzable and based on mathematical modeling. In the following, we refer to the definition of successful probability of communication and use beam sharing as the jointly designed technique to analyze the performance of cooperative detection. In our example, the radar and communication subsystems share the same spectrum and the main beam of radar is used for detection, while the sub-beam is used for communication. The single radar is subject to constraint of power, in which each radar occupy a power allocation coefficient expressed as $\beta$. A central radar exists for fusing detection information. Here, we define detection volume to represent the total detection area of multiple radars as the performance metric. Fig.1 illustrates the detection volume of cooperative detection as a function of $\beta$. The optimal fraction of power allocation between radar and communication has been obtained and the theoretical analysis provided a guideline for the collaboration of JRC with beam sharing.

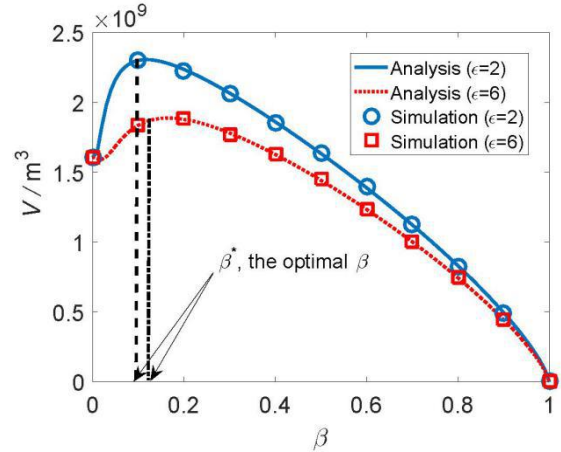

Fig.1 The volume of cooperative detection in multiple radars scenario, $V$, is calculated by JRC with beam sharing technique. Assuming that $\beta \in [0,1]$ is the fraction of the power allocated for communication. The transmission between the fusion center radar and other radars is successful only if the received $SINR_{com}$ at fusion center is larger than a threshold $\varepsilon$. The detection volume is only $1.6 \times 10^9$ m³ when the power is totally allocated to radar. As the coefficient $\beta$ increases, most of the power is allocated to the communication beam, which reduces the detection volume.

### 5.2 Radar-aided Wireless Networking

Nowadays, there is a strong demand for fast wireless networking in some high-mobility applications, such as networks of vehicles, aircrafts and robots. Traditional methods of wireless networking for machine-type communication (MTC) usually have a complex discovery process and the time consumption is huge. Undoubtedly, it reduces the networking efficiency and leads to a waste of resource. With extensive deployment of radars in these machines for environment awareness, the radar-aided wireless networking has the advantages of fast speed, low time consumption and discovery of multiple nodes at once. Therefore, the radar-aided wireless networking has attracted wide attention.

Neighbor discovery procedure is the first step of network initialization. It provides nodes information for MAC protocol, routing protocol and topology control protocol. Most algorithms discover neighbors without prior knowledge or utilize prior knowledge provided by other frequency bands. In recent years, some researches utilized the prior knowledge of radar to accelerate the traditional neighbor discovery [200], but they did not explore the node distribution, so the prior knowledge were not fully utilized.

In our research, we assume that every node possesses the dual-function of radar and communication. Radar is regard as an auxiliary to aid nodes networking, such as self-organization of unmanned aerial vehicle networks. The distribution law of nodes has been considered, and several neighbor discovery algorithms have been designed according to the utilization degree of prior knowledge, so as to provide the location information of neighbors for nodes and hence accelerate neighbor discovery. Fig.2 shows one of the neighbor discovery algorithm, named complete random algorithm based radar prior knowledge (CRA-RPK). Compared with the previous CRA, the proposed algorithms has significant advantage in time slots reduction.

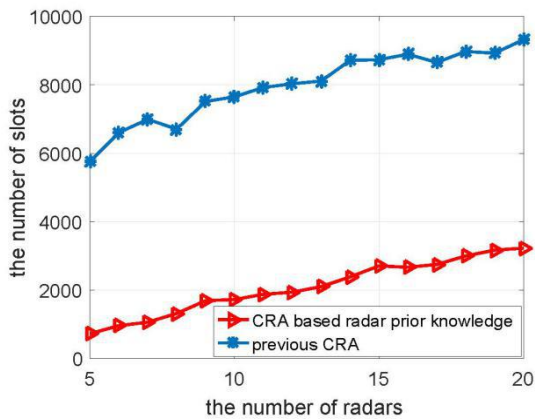

**Fig.2** X-axis represents the number of node in actual networks, and Y-axis represents the number of slots required within a completed neighbor discovery process. Based on the radar prior knowledge, the speed of neighbor discovery is about 6 times better than the previous CRA.

### 5.3 Radar-aided Channel Estimation and Beam Alignment

Because of the fast change of communication channel and the large scale of massive MIMO antenna, the channel estimation has been an intractable problem, which should be frequent enough to catch the fast change of Channel State Information (CSI). Thus, the cost of channel estimation and transmission of CSI will be large, leading to the huge problem of decrease of spectrum efficiency [201],[202]. JRC technique can achieve the reciprocity of channel estimation, because MIMO antenna and multi-carrier signal all can be utilized in radar and communication, and the propagation of radar and communication electromagnetic wave have similar propagation property [162]. The motion parameters and raw echo input on antenna array that radar function can provide useful information such as DoA and time delay for communication channel estimation.

In order to support the required high data rates in high-mobility scenarios, training-based beam sweeping has been regarded as the main approach at the present stage. However, it brings in the high overhead due to frequent training of beam. The concept of radar-aided beam alignment is an effective method to overcome this limitation, because radar can search the best beam pair and discover the optimal antenna alignment point quickly [203]. As a result, it can sharply save costs compared with traditional beam sweeping method. Furthermore, the DOA information can be directly obtained by using the directivity of radar.

### 5.4 Joint Resource Sharing

Another potential issue on the JRC techniques is that the resource allocation could become complex when a large number of resources (e.g. time, spectrum, beam, storage, calculation) need to be scheduled at the same time.

Radar with multi-standard, multi-band and detection capability differentiation drives the uniform methods of resource cognition, characterization, metric and partitioning. Therefore, it is necessary to develop a resource allocation model. Based on the resources parameters, we have proposed a resource sharing pool model as shown in Fig.3. In this model, multidimensional resources such as power, frequency, time slots, pilot code, multi-array RF, calculation, storage are constructed as a resource pool, and various tasks such as target detection, target tracking, and track establishment are constructed as a task pool. Considering the working bands, detection capability and multidimensional resource attributes are different from communication in radar system. The match between the diverse tasks and multidimensional resources can be realized based on the real time construction of resource pool. Furthermore, the multitasks have different requirements in terms of detection distance, target resolution, data parallel processing efficiency, multidimensional track data fusion, and update rate. That makes it necessary to use the intelligent algorithm to extract the characteristics of multitasks. Besides, we adopt the virtualization technology to shield the differences of physical resources, and then we can characterize and quantize the tasks and resources. Particularly, it is essentially realized that this pool model is suited for any resource allocation components of JRC systems.

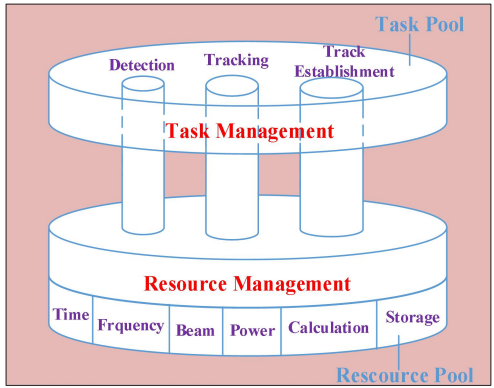

**Fig.3 The resource sharing pool model is grouped into two parts: task management and resource management.**

### 5.5 Multidimensional Signal Processing and Improved Information Theory

At present, JRC design is mainly oriented towards the demand of applications, the study of basic information theory is not profound enough. As is known to all, Shannon Theory lays the foundation of wireless communication. In the future, JRC should move to theoretical level rather than be based on the technical requirements, and further, build an unified improved information theory. That may includes: joint performance bounds, multidimensional signal processing and intelligent signal system.

Research on joint performance bounds is the foundation to establish the intelligent signal system. Bliss *et al* in [77] first proposed a novel approach named *estimation information rate* to evaluate the JRC performance bounds. This is of profound significance to the improved information theory.

Multidimensional signal processing is another important part for supporting the improved information theory. To carry out this part of work, multidimensional cooperative signal model and information transmission mechanism should be considered at first. The electromagnetic environment perception and collaborative inversion method of information should be considered as the second step. Finally, we should carry out the studies in intelligent control of electromagnetic spectrum.

Multidimensional signal has many resource dimensions such as time, space, frequency, coding and polarization. However, traditional signal model is limited to the mathematical representation of waveform. Therefore, constructing intelligent signal system is necessary. Considering the electromagnetic wave properties of multidimensional signals, the Maxwell's equations of multidimensional signals can further promote the construction of intelligent signal system.

### 5.6 Cost and Scalability Issues

With the development of JRC, the underlying cost is another important part that should be considered. The overall cost is significantly reduced from the current research results. We can look at it from two aspects: hardware costs and software costs. For hardware costs, the antenna co-use of radar and communication can reduce the number of hardware devices [23]. The jointly designed waveform enables a signal to have two functions, thus reducing the power consumption of transmission. Reference [96] proves that the JRC system can effectively save the transmitting power under the same code number. From the perspective of software, the current waveform design can rely on existing FPGA and MIMO antenna array to achieve the definition of software functions, and thus helping reduce the cost of research and production [204]. Furthermore, the SDN technology can also be applied to JRC to reduce costs [19]. In literature [162], analog antenna array is adopted for multibeam allocation, which has lower cost compared with digital antenna array. In addition, both the OFDM-based and LFM-based waveform design can be supported by existing algorithms so that reducing the cost of JRC design [90].

From the view of scalability, inserting some fixed bits into the joint waveform is an appropriate solution which can assist the communication decoding process and further improve the transmission performance without increasing the system complexity [205]. To ensure the need of signalling among nodes, appropriate multiple access protocol, suitable duplex technique and orderly resource scheduling are indispensable. Besides, the commercialization of 5G can enhance the expansion capability of JRC. In particular, 5G has lower delay and higher rate features, which can provide faster and more accurate information for radar-aided wireless networking. Furthermore, it can ensure the reliability of transmission signals. For example, the sensing base station will be a potential entity for JRC deployment in the background of 5G. Therefore, the scalability of JRC system can be increased significantly due to the large-scale deployment and wide coverage of 5G base stations. Secondly, using steerable arrays can also improve the scalability of JRC system. In literature [162], the steerable analog antenna arrays can be applied to those scenarios whose volume or power are limited, such as automotive vehicles. Lastly, road side units (RSUs) with sensing capability also have potential to help with the scalability of JRC system due to the large-scale number and easy deployment features.

## VI. CONCLUSIONS

Joint design of radar and communication is one of the potential orientation of the research on RF convergence system. In this paper, we have presented the basic concepts, characteristics and advantages of JRC technologies and introduced an updated state-of-the-art development according to four categories, namely, coexistence, cooperation, co-design and collaboration. The various applications that benefited from the JRC technologies and will gain great development in the future have been investigated in detail. Additionally, we have proposed a new concept of radar and communication operating with mutual benefits based on collaboration. Five new research orientations, namely, communication-aided multi-radar cooperative detection, radar-aided wireless networking, radar-aided channel estimation/beam alignment, resource sharing based pool model and the improved information theory, have been proposed, respectively. The cost and scalability issues also have been discussed in detail. We can foresee magnificent interaction opportunities for scholars to conduct researches on JRC technologies in the overall future society.

## REFERENCES

[1] HAN Y, EKICI E, KREMO H, et al. Spectrum Sharing Methods for the Coexistence of Multiple RF Systems: A Survey[J]. Ad Hoc Networks, 2016, 53: 53-78.

[2] CHEN Zhi, MA Xinying, ZHANG Bo, et al. A Survey on Terahertz Communications[J]. China Communications, 2019, 16(2): 1-35.

[3] LONG Teng, ZENG Tao, HU Cheng, et al. High Resolution Radar Real-time Signal and Information Processing[J]. China Communications, 2019, 16(2): 105-133.

[4] JACKSON C A, HOLLOWAY J R, POLLARD R, et al. Spectrally Efficient Radar Systems in the L and S Bands[C]// Proceedings of IET International Conference on Radar Systems. Edinburgh, UK: IET Press, 2007: 1-6.

[5] ZHANG Ping, YANG Xiaoli, CHEN Jianqiao, et al. A Survey of Testing for 5G: Solutions, Opportunities, and Challenges[J]. China Communications, 2019, 16(1): 69-85.

[6] WANG Lingfeng, MCGEEHAN J P, WILLIAMS C, et al. Application of Cooperative Sensing in Radar-Communications Coexistence[J]. IET Communications, 2008, 2(6): 856-868.

[7] PAUL B, CHIRIYATH A R, BLISS D W. Survey of RF Communications and Sensing Convergence Research[J]. IEEE Access, 2017, 5: 252-270.

[8] KHAWAR A, ABDEL-HADI A, CLANCY T C. MIMO Radar Waveform Design for Coexistence with Cellular Systems[C]// Proceedings of IEEE International Symposium on Dynamic Spectrum Access Networks. McLean, VA, USA: IEEE Press, 2014: 20-26.

[9] CAGER R, LAFLAME D, PARODE L. Orbiter Ku-Band Integrated Radar and Communications Subsystem[J]. IEEE Transactions on Communications, 1978, 26(11): 1604-1619.

[10] DARPA. The Defense Advanced Research Projects Agency (DARPA): Shared Spectrum Access for Radar and Communications (SSPARC)[R]. 2013.

[11] NIJSURE Y, CHEN Yifan, Said B, et al. Novel System Architecture and Waveform Design for Cognitive Radar Radio Networks[J]. IEEE Transactions on Vehicular Technology, 2012, 61(8): 3630-3642.

[12] STINCO P, GRECO M, GINI F, et al. Channel Parameters Estimation for Cognitive Radar Systems[C]// Proceedings of International Workshop on Cognitive Information Processing. Copenhagen, Denmark: IEEE Press, 2014: 1-6.

[13] QUAN Siji, QIAN Weiping, GUO Junhai, et al. Radar-Communication Integration: An Overview[C]// Proceedings of IEEE International Conference on Advanced Infocomm Technology. Fuzhou, China: IEEE Press, 2014: 98-103.

[14] HAMMED K, GHAURI S A, and QAMAR M S. Biological Inspired Stochastic Optimization Technique (PSO) for DOA and Amplitude Estimation of Antenna Arrays Signal Processing in Radar Communication System[J]. Journal of Sensors, 2016, 1-10.

[16] GAGLIONE D, CLEMENTE C, ILIOUDIS C V, et al. Fractional Fourier Transform Based Waveform for a Joint Radar-Communication System[C]// Proceedings of IEEE Radar Conference. Philadelphia, PA, USA: IEEE Press, 2016: 1-6.

[17] BICA M and KOIVUNEN V. Radar Waveform Optimization for Target Parameter Estimation in Cooperative Radar-Communications Systems[J]. IEEE Transactions on Aerospace and Electronic Systems, Early Access, 2018.

[18] LIU Fan, Zhou Longfei, MASOUROS C, et al. Toward Dual-functional Radar-Communication Systems: Optimal Waveform Design[J]. IEEE Transactions on Signal Processing, 2018, 66(16): 4264-4279.

[19] ROSSLER C W, ERTIN E, MOSES R L. A Software Defined Radar System for Joint Communication and Sensing[C]// Proceedings of IEEE Radar Conference (RadarConf). City, MO, USA: IEEE Press, 2011: 1050-1055.

[20] GUTIERREZ R M, HERSCHFELT A, YU Hanguang, et al. Joint Radar-Communications System Implementation Using Software Defined Radios: Feasibility and Results[C]// Proceedings of 51st Asilomar Conference on Signals, Systems, and Computers. Pacific Grove, CA, USA: IEEE Press, 2017:1127-1132.

[21] Rihan M, Huang Lei. Non-orthogonal Multiple Access Based Cooperative Spectrum Sharing between MIMO Radar and

MIMO Communication Systems[J]. Digital Signal Processing, 2018, 83: 107-117.

[22] MA O, CHIRIYATH A R, HERSCHFELT A, et al. Cooperative Radar and Communications Coexistence Using Reinforcement Learning[C]// Proceedings of 52nd Asilomar Conference on Signals, Systems, and Computers. Pacific Grove, CA, USA, USA: IEEE Press, 2018: 947-951.

[23] CHIRIYATH A R, PAUL B, BLISS D W. Radar-Communications Convergence: Coexistence, Cooperation, and Co-Design[J]. IEEE Transactions on Cognitive Communications and Networking, 2017, 3(1): 1-12.

[24] WU K, HAN Liang. Joint Wireless Communication and Radar Sensing Systems-State of the Art and Future Prospects[J]. IET Microwaves, Antennas & Propagation, 2013, 7(11):876-885.

[25] SCHARRENBROICH M, and ZATMAN M. Joint Radar Communications Resource Management[C]// Proceedings of IEEE Radar Conference (RadarConf16). Philadelphia, PA, USA: IEEE Press, 2016: 210-215.

[26] ZHOU Yifan., Zhou Huilin., Zhou Fuhui, et al. Resource Allocation for a Wireless Powered Integrated Radar and Communication system[J]. IEEE Wireless Communications Letters, 2019, 8(1): 253-256.

[27] KUMARI P, GONZALEZ-PRELCIC N, and ROBERT. W. H, Jr. Investigating the IEEE 802.11 ad Standard for Millimeter Wave Automotive Radar[C]// Proceedings of 82nd Vehicular Technology Conference (VTC2015-Fall). Boston, MA, USA : IEEE Press, 2015: 1-5.

[28] XIA Deping, ZHANG Yu, CAI Peixiang, et al. An Energy-Efficient Signal Detection Scheme for a Radar-Communication System Based on the Generalized Approximate Message-Passing Algorithm and Low-Precision Quantization[J]. IEEE Access, 2019, 7: 29065-29075.

[29] TRUNK G V. Advanced Multifunction RF System (AMRFS) Preliminary Design Considerations[R]. In NRL, Washington, DC, Formal Rep., 2001, 5300-01-9914.

[30] HUGHES P K, and CHOE J Y. Overview of Advanced Multifunction RF System (AMRFS)[C]// Proceedings of IEEE International Conference on Phased Array Systems and Technology. Dana Point, CA, USA: IEEE Press, 2000, 21-24.

[31] MOO P W, and DIFILIPPO D J. Overview of Naval Multifunction RF Systems[C]// Proceedings of 15th European Radar Conference (EuRAD). Dana Point, CA, USA: IEEE Press, 2018: 178-181.

[32] LONG B D. An Integrated Approach to Topside Design[C]// Proceedings of IEEE MTT-S International Microwave Symposium (IMS). Honolulu, HI, USA: IEEE Press, 2017: 326-329.

[33] FERRI* G, MUNAFO*, ANDREA, et al. Cooperative Robotic Networks for Underwater Surveillance: An Overview[J]. IET Radar, Sonar and Navigation, 2017, 11(12): 1740-1761.

[34] PAUL B, and BLISS D W. Constant Information Radar for Dynamic Shared Spectrum Access[C]// Proceedings of 49th Asilomar Conference on Signals, Systems and Computers. Pacific Grove, CA, USA: IEEE Press, 2015: 1374-1378.

[35] GUERCI J, GUERCI R, LACKPOUR A, et al. Joint Design and Operation of Shared Spectrum Access for Radar and Communications[C]// Proceedings of IEEE Radar Conference (RadarConf). Arlington, VA, USA: IEEE Press, 2015: 761–766.

[36] HAN You, EKICI E, KREMO H, et al. Optimal Spectrum Utilization in Joint Automotive Radar and Communication Networks[C]// Proceedings of 14th International Symposium on Modeling and Optimization in Mobile, Ad Hoc, and Wireless Networks (WiOpt). Tempe, AZ, USA: IEEE Press, 2016: pp. 1–8.

[37] GAMEIRO A, CASTANHEIRA D, SANSON J, et al. Research Challenges, Trends and Applications for Future Joint Radar Communications Systems[J]. Wireless Personal Communications, 2018, 100(1): 81-96 .

[38] DANIELS R C, YEH E R, and ROBERT. W. Heath, Jr. Forward Collision Vehicular Radar with IEEE 802.11: Feasibility Demonstration through Measurements[J]. IEEE Transactions on Vehicular Technology, 2018, 67(2): 1404-1416.

[39] KUMARI P, CHOI J, GONZALEZ-PRELCIC N, et al. IEEE 802.11ad-Based Radar: An Approach to Joint Vehicular Communication-Radar System[J]. IEEE Transactions on Vehicular Technology, 2018, 67(4): 3012-3027.

[40] KUMARI P, NGUYEN D H N, and ROBERT W Heath, Jr. Performance Trade-off in an Adaptive IEEE 802.11ad Waveform Design for a Joint Automotive Radar and Communication System[C]// Proceedings of International Conference on Acoustics, Speech, and Signal Processing (ICASSP). New Orleans, LA, USA: IEEE Press, 2017: 4281-4285.

[41] KUMARI P, ELTAYEB M E, and ROBERT W Heath, Jr. Sparsityaware Adaptive Beamforming Design for IEEE 802.11ad-Based Joint Communication-Radar[C]// Proceedings of IEEE Radar Conference (RadarConf18). Oklahoma City, OK, USA: IEEE Press, 2018: 923-928.

[42] CHOI J, VA V, GONZALEZ-PRELCIC N, et al. Millimeter-Wave Vehicular Communication to Support Massive Automotive Sensing[J]. IEEE Communications Magazine, 2016, 54(12): 160-167.

[43] REICHARDT L, STURM C, GRUNHAUPT F, et al. Demonstrating the Use of the IEEE 802.11p Car-to-Car Communication Standard for Automotive Radar[C]// Proceedings of European Conference on Antennas & Propagation. Prague, Czech Republic: IEEE Press, 2012: 1576-1580.

[44] DOKHANCHI S H, SHANKAR M R B, NIJSURE Y A, et al. Joint Automotive Radar-Communications Waveform Design[C]// Proceedings of International Symposium on Personal, Montreal, QC, Canada: IEEE Press, 2018: 1-7.

[45] SITY L, ZWICK T. Automotive MIMO OFDM radar: Subcarrier Allocation Techniques for Multiple-User Access and DOA Estimation[C]// Proceedings of 11th European Radar Conference, Rome, Italy: IEEE Press, 2014:153-156.

[46] Dokhanchi S H, SHANKAR M B, STIFTER T, et al. OFDM-Based Automotive Joint Radar Communication System[C]// Proceedings of IEEE Radar Conference (RadarConf18), Oklahoma City, OK, USA: IEEE Press, 2018: 902-907.

[47] HAN Liang, WU Ke. 24GHz Integrated Radio and Radar System Capable of Time-Agile Wireless Communication and Sensing[J]. IEEE Transactions on Microwave Theory & Techniques, 2012, 60(3): 619-631.

[48] HAN Liang, WU Ke. 24GHz Joint Radar and Radio System Capable of Time-Agile Wireless Sensing and Communication[C]// Proceedings of Microwave Symposium Digest, Baltimore, MD, USA: IEEE Press, 2011: 1-4.

[49] HAN Liang, WU Ke. Multifunctional Transceiver for Future Intelligent Transportation Systems[J]. IEEE Transactions on Microwave Theory and Techniques, 2011, 59(7):1879-1892.

[50] MOGHADDASI J, WU Ke. Improved Joint Radar-Radio (RadCom) Transceiver for Future Intelligent Transportation Platforms and Highly Mobile High-Speed Communication Systems[C]// Proceedings of Wireless Symposium, Beijing, China: IEEE Press, 2013: 1-4.

[51] MOGHADDASI J, WU Ke. Unified Radar-Communication (RadCom) Multi-Port Interferometer Transceiver[C]// Proceedings of IEEE Radar Conference, Nuremberg, Germany: IEEE Press, 2013: 1791-1794.

[52] MOGHADDASI J, WU Ke. Multifunctional Transceiver for Future Radar Sensing and Radio Communicating Data-Fusion Platform[J]. IEEE Access, 2016, 4: 818-828.

[53] KUMARI P, MAZHER K U, MEZGHANI A, et al. Low Resolution Sampling for Joint Millimeterwave MIMO Communication-Radar[C]// Proceedings of IEEE Statistical Signal Processing Workshop (SSP), Freiburg im Breisgau, Germany: IEEE Press, 2018: 193-197.

[54] STURM C, WIESBECK W. Waveform Design and Signal Processing Aspects for Fusion of Wireless Communications and Radar Sensing[J]. Proceedings of the IEEE, 2011, 99(7): 1236-1259.

[55] ABDULLAH R S A R, HASHIM F H, SALAH A A, et al. Experimental Investigation on Target Detection and Tracking in Passive Radar Using Long-Term Evolution Signal[J]. IET Radar Sonar & Navigation, 2015, 10(3): 577-585.

[56] REICHARDT L, MAURER J, FUGEN T, et al. Virtual Drive: A Complete V2X Communication and Radar System Simulator for Optimization of Multiple Antenna Systems[J]. Proceedings of the IEEE, 2011, 99(7): 1295-1310.

[57] SHI Xiufang, YANG Chaoqun, XIE Weige, et al. Anti-Drone System with Multiple Surveillance Technologies: Architecture, Implementation, and Challenges[J]. IEEE Communications Magazine, 2018, 56(4): 68-74.

[58] GUVENC I, KOOHIFAR F, SINGH S, et al. Detection, Tracking, and Interdiction for Amateur Drones[J]. IEEE Communications Magazine, 2018, 56(4): 75-81.

[59] SESAR-JU. U-Space Blueprint[R]. Jun. 2017.

[60] NASA. The Unmanned Air System Traffic Management (UTM) Directory by Unmanned Airspace.info[R]. Sep. 2017.

[61] IAI. IAI's ELTA Systems Unveils Next Generation Drone Guard Counter Unmanned Aircraft System (C-UAS)[R]. [Online]. Available:http://www.iai.co.il/2013/32981-49093-en/MediaRom.aspx.

[62] LEONARDO. LYRA 10 Surveillance Radar Product Information [Online]. Available: https://www.leonardocompany.com/en/-/lyra 10.

[63] MATTHIAS S, STROHMEIER M, LENDERS V, et al. Bring uo OpenSky: A Large-scale ADS-B Sensor Network for Research[C]// Proceedings of IPSN-14 Proceedings of the 13th International Symposium on Information Processing in Sensor Networks, Berlin, Germany: IEEE Press, 2014: 83-94.

[65] STROHMEIER M, SCHAFER M, LENDERS V, et al. Realities and Challenges of Nextgen Air Traffic Management: the Case of ADS-B[J]. IEEE Communications Magazine, 2014, 52(5): 111-118.

[66] FILIP A, and SHUTIN D. Cramer-Rao Bounds for L-Band Digital Aeronautical Communication System Type 1 Based Passive Multiple-Input Multiple-Output Radar[J]. IET Radar, Sonar & Navigation, 2016, 10(2) : 348-358.

[68] MCCORMICK P M, RAVENSCROFT B, BLUNT S D, et al. Simultaneous Radar and Communication Emissions from a Common Aperture, Part II: Experimentation[C]// Proceedings of IEEE Radar Conference (RadarConf17), Seattle, WA, USA: IEEE Press, 2017: 1697-1702.

[69] GARMATYUK D, SCHUERGER J, KAUFFMAN K, et al. Wideband OFDM System for Radar and Communications[C]// Proceedings of IEEE Radar Conference (RadarConf), Pasadena, CA, USA: IEEE Press, 2009: 1-6.

[70] LABIB M, MAROJEVIC V, MARTONE A F, et al. Coexistence Between Communication and Radar Systems-A Survey[J]. Radio Science, 2017, 2017(362): 74-82.

[71] HAN Jinpeng, WANG Ben, WANG Weidong, et al. Analysis for the BER of LTE System with the Interference from Radar[C]// Proceedings of IET International Conference on Communication Technology and Application (ICCTA 2011), Beijing, China: IEEE Press, 2011: 452-456.

[72] NARTASILPA N, TUNINETTI D, DEVROYE N, et al. Let's Share CommRad: Effect of Radar Interference on an Uncoded Data Communication System[C]// Proceedings of IEEE Radar Conference (RadarConf16), Philadelphia, PA, USA: IEEE Press, 2016: 1-5.

[73] GHORBANZADEH M, VISOTSKY E, MOORUT P, et al. Radar Inband and Out-of-Band Interference into LTE Macro and Small Cell Uplinks in the 3.5 GHz Band[C]// Proceedings of IEEE Wireless Communications and Networking Conference (WCNC), New Orleans, LA, USA: IEEE Press, 2015: 1829-1834.

[74] REED J H, CLEGG A W, PADAKI A V, et al. On the Co-existence of TD-LTE and Radar over 3.5 GHz Band: An Experimental Study[J]. IEEE Wireless Communication Letters, 2016, 5(4): 368-371.

[75] BELL M R, DEVROYE N, ERRICOLO D, et al. Results on Spectrum Sharing Between a Radar and a Communications System[C]// Proceedings of International Conference on Electromagnetics in Advanced Applications (ICEAA), Palm Beach, Netherlands Antilles: IEEE Press, 2014: 826-829.

[76] CORDILL B D, SEGUIN S A, COHEN L. Electromagnetic Interference to Radar Receivers Due to In-band OFDM Communications Systems[C]// Proceedings of IEEE International Symposium on Electromagnetic Compatibility, Denver, CO, USA: IEEE Press, 2013: 72-75.

[77] LIU Wei, FANG Jian, TAN Haifeng, et al. Coexistence Studies for TD-LTE with Radar System in the Band 2300-2400 MHz[C]// Proceedings of International Conference on Communications, Chengdu, China: IEEE Press, 2010: 49-53.

[78] KHAWAR A, ABDEL-HADI A, CLANCY C T. A Mathematical Analysis of LTE Interference on the Performance of S-band Military Radar Systems[C]// Proceedings of Wireless Telecommunications Symposium, Washington, DC, USA: IEEE Press, 2014: 1-8.

[79] ZHENG Le, LOPS M, WANG Xiaodong, et al. Joint Design of Co-existing Communication System and Pulsed Radar[C]// Proceedings of IEEE International Workshop on Computational Advances in Multi-sensor Adaptive Processing, Curacao, Netherlands Antilles: IEEE Press, 2018: 139-154.

[80] CHIRIYATH A R, PAUL B, JACYNA G M, et al. Inner Bounds on Performance of Radar and Communications Co-Existence. IEEE Transactions on Signal Processing, 2015, 64(2): 464-474.

[81] CUI Yuanhao, Koivunen V, and JING Xiaojun. Interference Alignment Based Precoder-Decoder Design for Radar Communication Co-existence[C]// Proceedings of 51st Asilomar Conference on Signals, Systems, and Computers, Pacific Grove, CA, USA: IEEE Press, 2017: 1290-1295.

[83] ZHENG Le, LOPS M, WANG Xiaodong. Adaptive Interference Removal for Un-coordinated Radar/Communication Co-existence[J]. IEEE Journal of Selected Topics in Signal Processing, 2017, 12(1): 45-60.

[84] SHAN Chengzhao, MA Yongkui, ZHAO Honglin, et al. Joint Radar-Communications Design Based on Time Modulated Array[J]. Digital Signal Processing, 2018, 82: 43-53.

[85] REN Ping, MUNARI A, PETROVA M. Performance Tradeoffs of Joint Radar-Communication Networks[J]. IEEE Wireless Communications Letters, 2019, 8(1): 165-168.

[86] DENG Hai, HIMED B. Interference Mitigation Processing for Spectrum-Sharing Between Radar and Wireless Communications Systems[J]. IEEE Transactions on Aerospace and Electronic Systems, 2013, 49(3):1911-1919.

[87] LI Bo, KUMAR H, and PETROPULU A P. A Joint Design Approach for Spectrum Sharing Between Radar and Communication Systems[C]// Proceedings of IEEE International Conference on Acoustics, Speech and Signal Processing Proceedings, Shanghai, China: IEEE Press, 2016: 3306-3310.

[88] LI Bo, and PETROPULU A P. Spectrum Sharing Between Matrix Completion Based MIMO Radars and a MIMO Communication System[C]// Proceedings of IEEE International Conference on Acoustics, Speech and Signal Processing (ICASSP), Brisbane, QLD, Australia: IEEE Press, 2015: 2444-2448.

[89] LI Bo, PETROPULU A P. Joint Transmit Designs for Co-existence of MIMO Wireless Communications and Sparse Sensing Radars in Clutter[J]. IEEE Transactions on Aerospace and Electronic Systems, 2017, 53(6): 2846-2864.

[90] LI Bo, PETROPULU A P, TRAPPE W. Optimum Co-Design for Spectrum Sharing Between Matrix Completion Based MIMO Radars and a MIMO Communication System[J]. IEEE Transactions on Signal Processing, 2016, 64(17): 4562-4575.

[91] QIAN Junhui, Lops M, ZHENG Le, et al. Joint Design for Co-existence of MIMO Radar and MIMO Communication System[C]// Proceedings of 51st Asilomar Conference on Signals, Systems, and Computers, Pacific Grove, CA, USA: IEEE Press, 2017: 568-572.

[92] QIAN Junhui, Lops M, ZHENG Le, et al. Joint System Design for Co-existence of MIMO Radar and MIMO Communication[J]. IEEE Transactions on Signal Processing, 2018, 66(13): 3504-3519.

[93] QIAN Junhui, HE Zishu, HUANG Nuo, et al. Transmit Designs for Spectral Coexistence of MIMO Radar and MIMO Communication System[J]. IEEE Transactions on Circuits and Systems II: Express Briefs, 2018, 65(12): 2072-2076.

[94] LIU Fan, MASOUROS C, LI A, et al. MU-MIMO Communications with MIMO Radar: From Co-existence to Joint Transmission[J]. IEEE Transactions on Wireless Communications, 2017, 17(4) :2755-2770.

[95] ARIK M. and AKAN O B. Realizing Joint Radar Communications in Coherent MIMO Radars[C]// Proceedings of Physical Communication, 2019, 32: 145-159.

[96] LI Bo, PETROPULU A P. MIMO Radar and Communication Spectrum Sharing with Clutter Mitigation[C]// Proceedings of IEEE Radar Conference (RadarConf16), Philadelphia, PA, USA: IEEE Press, 2016: 1-5.

[97] ZHENG Le, LOPS M, WANG Xiaodong, et al. Joint Design of Overlaid Communication Systems and Pulsed Radars[J]. IEEE Transactions on Signal Processing, 2018, 66(1): 139-154.

[98] BISWAS S, SINGH K, TAGHIZADEH O, et al. Coexistence of MIMO Radar and FD MIMO Cellular Systems with QoS Considerations[J]. IEEE Transactions on Wireless Communications, 2018, 17(11): 7281-7294.

[99] BISWAS S, SINGH K, TAGHIZADEH O, et al. QoS-Based Robust Transceiver Design for Coexistence of MIMO Radar and FD MU-MIMO Cellular System[C]// Proceedings of IEEE Global Communications Conference (GLOBECOM), Abu Dhabi, United Arab Emirates, United Arab Emirates: IEEE Press, 2018: 1-7.

[100] LIU Fan, MASOUROS C, LI A, et al. MIMO Radar and Cellular Coexistence: A Power-Efficient Approach Enabled by Interference Exploitation[J]. IEEE Transactions on Signal Processing, 2018, 66(14): 3681-3695.

[101] HONG J H, CHOI S W, KIM C S, et al. Interference Measurement Between 3.5 GHz 5G System and Radar[C]// Proceedings of International Conference on Information and Communication Technology Convergence (ICTC), Jeju, South Korea: IEEE Press, 2018: 1539-1541.

[102] LANGMAN A, HAZARIKA O, MISHRA A K, et al. White RHINO: A Low Cost Whitespace Communications and Radar Hardware Platform[C]// Proceedings of 23rd International Conference Radioelektronika (RADIOELEKTRONIKA), Pardubice, Czech Republic: IEEE Press, 2013: 240-244.

[103] MISHRA A K, INGGS M. White Space Symbiotic Radar: A New Scheme for Coexistence of Radio Communications and Radar[C]// Proceedings of IEEE Radar Conference (RadarConf15), Johannesburg, South Africa: IEEE Press, 2015: 56-60.

[104] RICHMOND C D, BASU P. Bayesian Framework and Radar: On Misspecified Bounds and Radar-Communication Cooperation[C]// Proceedings of IEEE Statistical Signal Processing Workshop (SSP), Palma de Mallorca, Spain: IEEE Press, 2016: 1-4.

[105] RICHMOND C D, BASU P, LEARNED R E, et al. Performance Bounds on Cooperative Radar and Communication Systems

[107] CHALISE B K, AMIN M, HIMED B. Performance Tradeoff in a Unified Passive Radar and Communications System[J]. IEEE Signal Processing Letters, 2017, 24(9): 1275-1279.

operation[C]// Proceedings of IEEE Radar Conference (RadarConf16), Philadelphia, PA, USA: IEEE Press, 2016: 1-6.

[108] CHALISE B K, and HIMED B. Performance Tradeoff in a Unified Multi-static Passive Radar and Communication System[C]// Proceedings of IEEE Radar Conference (RadarConf18), Oklahoma City, OK, USA: IEEE Press, 2018: 653-658.

[109] SHI Chenguang, WANG Fei, SANA S, et al. Optimal Power Allocation Strategy in a Joint Bistatic Radar and Communication System Based on Low Probability of Intercept[J]. Sensors, 2017, 17(12):2731-2748.

[110] HARPER A, REED J, ODOM J, et al. Performance of a Linear-Detector Joint Radar-Communication System in Doubly-Selective Channels[J]. IEEE Transactions on Aerospace and Electronic Systems, 2017, 53(2): 703-715.

[111] HE Qian, WANG Zhen, HU Jianbin, et al. Performance Gains From Cooperative MIMO Radar and MIMO Communication Systems[J]. IEEE Signal Processing Letters, 2019, 26(1): 194-198.

[112] CHIRIYATH A R, PAUL B, BLISS D W. Simultaneous Radar Detection and Communications Performance with Clutter Mitigation[C]// Proceedings of IEEE Radar Conference (RadarConf17), Seattle, WA, USA: IEEE Press, 2017: 279-284.

[113] HASSANIEN A, AMIN M G, ZHANG Y D, et al. A Dual Function Radar-Communications System Using Sidelobe Control and Waveform Diversity[C]// Proceedings of IEEE RadarConference (RadarConf15), Arlington, VA, USA: IEEE Press, 2015: 1260-1263.

[114] HASSANIEN A, AMIN M G, ZHANG Y D, et al. Dual-FunctionRadar-Communications:Information Embedding Using Sidelobe Control and Waveform Diversity[J]. IEEE Transactions on Signal Processing, 2016, 64(8): 2168-2181.

[115] HASSANIEN A, ABOUTANIOS E, AMIN M G, et al. A Dual-Function MIMO Radar-Communication System via Waveform Permutation[J]. Digital Signal Processing, 2018, 83: 118-128.

[116] NUSENU S Y, and WANG Wenqin. Dual-Fnction FDA MIMO Radar-Communications System Employing Costas Signal Waveforms[C]// Proceedings of IEEE Radar Conference (RadarConf18), Oklahoma City, OK, USA: IEEE Press, 2018: 33-38.

[117] JI Shilong, ChEN Hui, HU Quan, et al. A Dual-Function Radar-Communication System Using FDA[C]// Proceedings of IEEE Radar Conference (RadarConf18), Oklahoma City, OK, USA: IEEE Press, 2018: 224-229.

[118] NUSENU S Y, WANG Wenqin, BASIT A. Time-Modulated FD-MIMO Array for Integrated Radar and Communication Systems[J]. IEEE Antennas & Wireless Propagation Letters, 2018, 17(6): 1015-1019.

[119] ANTONIK P, WICKS M C, GRIFFITHS H D, et al. Frequency Diverse Array Radars[C]// Proceedings of IEEE Conference on Radar (RadarConf), Verona, NY, USA, USA: IEEE Press, 2006: 3.

[120] NUSENU S Y, SHAO Huaizong, PAN Ye, et al. Dual-Function Radar-Communication System Design Via Sidelobe Manipulation Based On FDA Butler Matrix[J]. IEEE Antennas and Wireless Propagation Letters, 2019, 18(3): 452-456.

[121] NOSRATI H, ABOUTANIOS E, SMITH D. Array Partitioning for Multi-task Operation in Dual Function MIMO Systems[J]. Digital Signal Processing, 2018, 82: 106-117.

[122] WANG Xiangrong, HASSANIEN A, AMIN M. G. Sparse Transmit Array Design for Dual-Function Radar Communications by Antenna Selection[J]. Digital Signal Processing, 2018, 83: 223-234.

[123] HASSANIEN A, AMIN M G, ZHANG Yimin D. Computationally Efficient Beampattern Synthesis for Dual-Function Radar-Communications[J]. Proceedings of SPIE, 2016, 9829: 98290L-1-98290L-8.

[124] HASSANIEN A, AMIN M G. Efficient Sidelobe ASK Based Dual-Function Radar-Communications[J]. Proceedings of SPIE, 2016, 9829: 98290K-1-98290K-10.

[125] ARIK M, and AKAN O B. Utilizing Sidelobe ASK Based Joint Radar-Communication System under Fading[C]// Proceedings of IEEE Military Communications Conference (MILCOM), Baltimore, MD, USA: IEEE Press, 2017: 623-628.

[126] HASSANIEN A, AMIN M G, ZHANG Yimin D, et al. Phase-Modulation Based Dual-Function Radar-Communications [J]. IET Radar, Sonar & Navigation, 2017, 10(8): 1411-1421.

[127] TEDESSO T W, and ROMERO R. Code Shift Keying Based Joint Radar and Communications for EMCON Applications[J]. Digital Signal Processing, 2018, 80: 48-56.

[128] AHMED A, ZHANG Yimin D, GU Yujie. Dual-Function Radar-Communications Using QAM-Based Sidelobe Modulation[J]. Digital Signal Processing, 2018, 82: 166-174.

[129] YAO Yu, ZHAO Junhui, WU Lenan. Adaptive Waveform Design for MIMO Radar-Communication Transceiver[J]. Sensors, 2018, 18(6): 1-16.

[130] GAGLIONE D, CLEMENTE C, ILIOUDIS C, et al. Waveform Design for Communicating Radar Systems Using Fractional Fourier Transform[J]. Digital Signal Processing, 2018, 80: 57-69.

[131] GU Yabin, ZHANG Linrang, ZHOU Yu, et al. Waveform Design for Integrated Radar and Communication System with Orthogonal Frequency Modulation[J]. Digital Signal Processing, 2018, 83: 129-138.

[132] ABDELHADI A, CLANCY T C. Network MIMO with Partial Cooperation between Radar and Cellular Systems[C]// Proceedings of International Conference on Computing, Networking and Communications (ICNC), Kauai, HI, USA: IEEE Press, 2015: 1-5.

[133] LIU Yongjun, LIAO Guisheng, XU Jingwei, et al. Transmit Power Adaptation for Orthogonal Frequency Division Multiplexing Integrated Radar and Communication Systems [J]. Journal of Applied Remote Sensing, 2017, 11(3): 1.

[134] GUERCI J R, GUERCI R M. RAST: Radar as a Subscriber Technology for Wireless Spectrum Cohabitation[C]// Proceedings of IEEE Radar Conference (RadarConf14), Cincinnati, OH, USA: IEEE Press, 2014: 1130-1134.

[135] SHAJAIAH H J, KHAWAR A, ABDEL-HADI A, et al. Resource Allocation with Carrier Aggregation in LTE Advanced Cellular System Sharing Spectrum with S-band Radar[C]// Proceedings of IEEE International Symposium on Dynamic Spectrum Access Networks (DYSPAN), McLean, VA, USA: IEEE Press, 2014: 34-37.

[136] ROBERTON M, BROWN E. Integrated Radar and Communications Based on Chirped Spread-Spectrum Techniques[C]// Proceedings of International Microwave Symposium Digest, Philadelphia, PA, USA, USA: IEEE Press, 2003: 611-614.

[137] XU Shaojian, CHEN Bing, ZHANG Ping. Radar-Communication Integration Based on DSSS Techniques[C]// Proceedings of International Conference on Signal Processing, Beijing, China: IEEE Press, 2007: 1-4.

[138] XU Shaojian, CHEN Yan, ZHANG Ping. Integrated Radar and Communication Based on DS-UWB[C]// Ultrawideband & Ultrashort Impulse Signals, the Third International Conference, Sevastopol, Ukraine: IEEE Press, 2007: 142-144.

[139] XIE Yanan, TAO Ran, WANG Teng. Method of Waveform Design for Radar and Communication Integrated System Based on CSS[C]// Proceedings of First International Conference on Instrumentation, Beijing, China: IEEE Press, 2011: 737-739.

[140] CHEN Xingbo, WANG Xiaomo, XU Shanfeng, et al. A Novel Radar Waveform Compatible with Communication[C]// Proceedings of International Conference on Computational Problem-solving, Chengdu, China: IEEE Press, 2011: 177-181.

[141] ZHANG Yu, LI Qingyu, HUANG Ling, et al. Waveform Design for Joint Radar-Communication with Nonideal Power Amplifier and Outband Interference[C]// Proceedings of Wireless Communications & Networking Conference (WCNC), San Francisco, CA, USA: IEEE Press, 2017: 1-6.

[142] ZHANG Zhiping, NOWAK M J, WICKS M, et al. Bio-Inspired RF Steganography via Linear Chirp Radar Signals[J]. IEEE Communications Magazine, 2016, 54(6): 82-86.

[143] NOWAK M J, WICKS M, ZHANG Zhiping, et al. Co-Designed Radar-Communication Using Linear Frequency Modulation Waveform[J]. IEEE Aerospace and Electronic Systems Magazine, 2016, 31(10): 28-35.

[144] ZHANG Zhiping, QU Yang, WU Zhiqiang, et al. RF Steganography via LFM Chirp Radar Signals[J]. IEEE Transactions on Aerospace and Electronic Systems, 2018, 54(3): 1221-1236.

[145] ELLINGER J, ZHANG Zhiping, WICKS M, et al. Dual-Use Multi-Carrier Waveform for Radar Detection and Communication[J]. IEEE Transactions on Aerospace and Electronic Systems, 2018, 54(3): 1265-1278.

[146] HOWARD S D, CALDERBANK A R, MORAN W. The Finite Heisenberg-Weyl Groups in Radar and Communications[J]. EURASIP Journal on Advances in Signal Processing, 2006(1): 1-13.

[147] PEZESHKI A, CALDERBANK A R, MORAN W, et al. Doppler Resilient Golay Complementary Waveforms[J]. IEEE Transactions on Information Theory, 2008, 54(9): 4254-4266.

[148] ZHANG Wenyi, VEDANTAM S, MITRA U. Joint Transmission and State Estimation: A Constrained Channel Coding Approach[J]. IEEE Transactions on Information Theory, 2011, 57(10): 7084-7095.

[149] LI Xiaobai, YANG Ruijuan, ZHANG Zunquan, et al. Research of Constructing Method of Complete Complementary Sequence in Integrated Radar and Communication[C]// Proceedings of IEEE 11th International Conference on Signal Processing (ICSP), Beijing, China: IEEE Press, 2012: 1729-1732.

[150] KELLETT D, GARMATYUK D, MORTON Y T J, et al. Radar Communications via Random Sequence Encoding[C]// Proceedings of 18th International Radar Symposium (IRS), Prague, Czech Republic: IEEE Press, 2017: 1-9.

[151] STURM C, ZWICK T, WIESBECK W. An OFDM System Concept for Joint Radar and Communications Operations[C]// IEEE Vehicular Technology Conference (VTC Spring), Barcelona, Spain: IEEE Press, 2009: 3339-3348.

[152] STURM C, ZWICK T, WIESBECK W, et al. Performance Verification of Symbol-Based OFDM Radar Processing[C]// Proceedings of IEEE Radar Conference (RadarConf), Washington, DC, USA: IEEE Press, 2010: 60-63.

[153] SIT Y L, REICHARDT L, STURM C, et al. Extension of the OFDM Joint Radar-Communication System for a Multipath, Multiuser Scenario[C]// Proceedings of IEEE Radar Conference (RadarConf), Kansas City, MO, USA: IEEE Press, 2011: 718-723.

[154] SIT Y L, STURM C, ZWICK T. Interference Cancellation for Dynamic Range Improvement in an OFDM Joint Radar and Communication System[C]// Proceedings of 8th European Radar Conference, Manchester, UK: IEEE Press, 2011: 333-336.

[155] SIT Y L, STURM C, ZWICK T. One-Stage Selective Interference Cancellation for the OFDM Joint Radar-Communication System[C]// Proceedings of 7th German Microwave Conference, Ilmenau, Germany: IEEE Press, 2012: 1-4.

[157] BRAUN M, TANBOURGI R, JONDRAL F K. Co-channel Interference Limitations of OFDM Communication-Radar Networks[J]. EURASIP Journal on Wireless Communications and Networking, 2013, 2013(1): 207-223.

[158] SIT Y L, STURM C, ZWICK T. Doppler Estimation in an OFDM Joint Radar and Communication System[C]// Proceedings of 2011 German Microwave Conference, Darmstadt, Germany: IEEE Press, 2011: 1-4.

[159] GOGINENI S, RANGASWAMY M, NEHORAI A. Multi-modal OFDM Waveform Design[C]// Proceedings of IEEE Radar Conference (RadarConf13), Ottawa, ON, Canada: IEEE Press, 2013: 1-5.

[160] BICA M, HUANG K W, KOIVUNEN V, et al. Mutual Information Based Radar Waveform Design for Joint Radar and Cellular Communication Systems[C]// Proceedings of IEEE International Conference on Acoustics, Speech and Signal Processing (ICASSP), Shanghai, China: IEEE Press, 2016: 3671-3675.

[161] LIU Yongjun, LIAO Guisheng, XU Jingwei, et al. Adaptive OFDM Integrated Radar and Communications Waveform Design Based on Information Theory[J]. IEEE Communications Letters, 2017, 21(10): 2174-2177.

[162] SHI Chenguang, SALOUS S, WANG Fei. Low Probability of Intercept-Based Adaptive Radar Waveform Optimization in Signal-Dependent Clutter for Joint Radar and Cellular

Communication Systems[J]. EURASIP Journal on Advances in Signal Processing, 2016, 2016(1): 1-13.

[163] SHI Chenguang, WANG Fei, MATHINI S, *et al*. Low Probability of Intercept Based Multicarrier Radar Jamming Power Allocation for Joint Radar and Wireless Communications Systems[J]. IET Radar Sonar & Navigation, 2017, 11(5): 802-811.

[164] SHI Chenguang, WANG Fei, SALOUS S, *et al*. Low Probability of Intercept-Based Optimal OFDM Waveform Design Strategy for an Integrated Radar and Communications System[J]. IEEE Access, 2018, 6: 57689-57699.

[165] TIAN Xuanxuan, and SONGT Z. On Radar and Communication Integrated System Using OFDM Signal[C]// Proceedings of IEEE Radar Conference (RadarConf17), Seattle, WA, USA: IEEE Press, 2017: 318-323.

[166] ZHANG Yu, LI Qingyu, HUANG Ling, *et al*. Waveform Design for Joint Radar-Communication System with Multi-user Based on MIMO Radar[C]// Proceedings of IEEE Radar Conference (RadarConf17), Seattle, WA, USA: IEEE Press, 2017: 415-418.

[167] ZHANG J A, CANTONI A, HUANG Xiaojing, et al. Joint Communications and Sensing Using Two Steerable Analog Antenna Arrays[C]// Proceedings of IEEE 85th Vehicular Technology Conference (VTC-Spring), Sydney, NSW, Australia: IEEE Press, 2017: 1-5.

[168] ZHANG J A, HUANG Xiaojing, GUO Yingjie, *et al*. Multibeam for Joint Communication and Sensing Using Steerable Analog Antenna Arrays[J]. IEEE Transactions on Vehicular Technology, 2019, 68(1): 671-685.

[169] ZHANG Wenshu, and YANG Liuqing. Communications-Inspired Sensing: a Case Study on Waveform Design.[J]. IEEE Transactions on Signal Processing, 2010, 58(2): 792-803.

[170] SINGH K, BISWAS S, GUPTA A, *et al*. Joint Power Allocation and Beamforming Design for Full-Duplex MIMO Cellular Systems with Spectrum Sharing Radar[C]// Proceedings of NASA/ESA Conference on Adaptive Hardware and Systems (AHS), Pasadena, CA, USA: IEEE Press, 2017: 93-100.

[171] RIHAN M, and HUANG Lei. Optimum Co-design of Spectrum Sharing between MIMO Radar and MIMO Communication Systems: An Interference Alignment Approach[J]. IEEE Transactions on Vehicular Technology, 2018, 67(12): 11667-11680.

[173] SABHARWAL A, SCHNITER P, GUO Dongning, *et al*. In-Band Full-Duplex Wireless: Challenges and Opportunities[J]. IEEE Journal on Selected Areas in Communications, 2014, 32(9): 1637-1652.

[174] BAI Ting, HU Hanying, and SONG Xiyu. OFDM MIMO Radar Waveform Design with High Range Resolution and Low Sidelobe Level[C]// Proceedings of IEEE 17th International Conference on Communication Technology (ICCT), Chengdu, China: IEEE Press, 2017: 1065-1069.

[175] LIU Yongjun, LIAO Guisheng, YANG Zhiwei. Range and Angle Estimation for MIMO-OFDM Integrated Radar and Communication Systems[C]// Proceedings of CIE International Conference on Radar (RADAR), Guangzhou, China: IEEE Press, 2016: 1-4.

[176] LIU Yongjun, LIAO Guisheng, YANG Zhiwei, *et al*. Design of Integrated Radar and Communication System Based on MIMO-OFDM Waveform[J]. Journal of Systems Engineering and Electronics, 2017, 28(4): 669-680.

[178] SIT Y L, and ZWICK T. MIMO OFDM Radar with Communication and Interference Cancellation Features[C]// Proceedings of IEEE Radar Conference (RadarConf14), Cincinnati, OH, USA: IEEE Press, 2014: 265-268.

[179] SIT Y L, NUSS B, and ZWICK T. On Mutual Interference Cancellation in a MIMO OFDM Multi-User Radar-Communication Network[J]. IEEE Transactions on Vehicular Technology, 2018, 67(4): 3339-3348.

[180] SIT Y L, NUSS B, BASAK S, *et al*. Demonstration of Interference Cancellation in a Multiple-User Access OFDM MIMO Radar-Communication Network Using USRPs[C]// Proceedings of IEEE MTT-S International Conference on Microwaves for Intelligent Mobility, San Diego, CA, USA: IEEE Press, 2016: 1-4.

[181] ZHANG Jianshu, PODKURKOV I, HAARDT M, et al. Efficient Multidimensional Parameter Estimation for Joint Wideband Radar and Communication Systems Based on OFDM[C]// Proceedings of IEEE International Conference on Acoustics, Speech and Signal Processing (ICASSP), New Orleans, LA, USA: IEEE Press, 2017: 3096-3100.

[182] ZHANG Yuxi, WANG Zhanchao, and WANG Jun. Integrated Radar Signal Processing Using FPGA Dynamic Reconfiguration [C]// Proceedings of CIE International Conference on Radar (RADAR), Guang zhou, China: IEEE Press, 2016: 1-4.

[183] ZHAO Licheng, and PALOMAR D P. Maximin Joint Optimization of Transmitting Code and Receiving Filter in Radar and Communications[J]. IEEE Transactions on Signal Processing, 2017, 65(4): 850-863.

[184] HERSCHFELT A, and Bliss D W. Joint Radar-Communications Waveform Multiple Access and Synthetic Aperture Radar Receiver[C]// Proceedings of 51st Asilomar Conference on Signals, Systems, and Computers, Pacific Grove, CA, USA: IEEE Press, 2017: 69-74.

[185] CABRERA J B, BASU P, WATSON W D, et al. Optimal Radar-Communications Spectral Maneuvering for TDOA-Based Tracking [C]// Proceedings of 52nd Asilomar Conference on Signals, Systems, and Computers, Pacific Grove, CA, USA: IEEE Press, 2018: 414-416.

[186] HERSCHFELT A, and BLISS D W. Spectrum Management and Advanced Receiver Techniques (SMART): Joint Radar-Communications Network Performance[C]// Proceedings of IEEE Radar Conference, Oklahoma City, OK, USA: IEEE Press, 2018: 1078-1083.

[187] HE Qian, HU Jianbin, BLUM R, *et al*. Generalized Cramer-Rao Bound for Joint Estimation of Target Position and Velocity for Active and Passive Radar Networks[J]. IEEE Transactions on Signal Processing, 2016, 64(8): 2078-2089.

[188] SHI Chenguang, WANG Fei, SALOUS S, *et al*. Performance Analysis for Joint Target Parameter Estimation in UMTS-Based Passive Multistatic Radar with Antenna Arrays Using Modified Cramer-Rao Lower Bounds[J]. Sensors, 2017, 17(10): 1-24.

[189] WU Yonggang, HE Qian, HU Jianbin, *et al*. Demonstrating Significant Passive Radar Performance Increase through Using Known Communication Signal Format[C]// Proceedings of Fifty-First Asilomar Conference on Signals, Systems, and Computers, Pacific Grove, CA, USA: IEEE Press, 2017: 75-79.

[190] BLISS D W. Cooperative Radar and Communications Signaling: The Estimation and Information Theory Odd Couple[C]// Proceedings of IEEE Radar Conference (RadarConf14), Cincinnati, OH, USA: IEEE Press, 2014: 50-55.

[191] PAUL B, BLISS D W. Extending Joint Radar-Communications Bounds for FMCW Radar with Doppler Estimation[C]// Proceedings of IEEE Radar Conference (RadarConf15), Arlington, VA, USA: IEEE Press, 2015: 89-94.

[192] PAUL B, CHIRIYATH A R, and BLISS D W. Joint Communications and Radar Performance Bounds under Continuous Wave-form Optimization: The Waveform Awakens[C]// Proceedings of IEEE Radar Conference (RadarConf16), Philadelphia, PA, USA: IEEE Press, 2016: 1-6.

[194] RONG Yu, CHIRIYATH A R, and BLISS D W. Multiple-Antenna Multiple-Access Joint Radar and Communications Systems Performance Bounds[C]// Proceedings of Asilomar Conference on Signals, Pacific Grove, CA, USA: IEEE Press, 2018: 1296-1300.

[195] RONG Yu, CHIRIYATH A R, and BLISS D W. MIMO Radar and Communications Spectrum Sharing: A Multiple-Access Perspective [C]// Proceedings of IEEE 10th Sensor Array and Multichannel Signal Processing Workshop (SAM), Sheffield, UK: IEEE Press, 2018: 272-276.

[196] CHIRIYATH A R, PAUL B, and BLISS D W. Joint Radar-Communications Information Bounds with Clutter: The Phase Noise Menace[C]// Proceedings of IEEE Radar Conference (RadarConf16), Philadelphia, PA, USA: IEEE Press, 2016: 1-6.

[197] JEONG S, KHALILI S, SIMEONE O, et al. Multistatic Cloud Radar Systems: Joint Sensing and Communication Design[J]. Transactions on Emerging Telecommunications Technologies, 2016, 27(5): 716-730.

[198] KILANI M B, GAGNON G, GAGNON F. Multistatic Radar Placement Optimization for Cooperative Radar-Communication Systems[J]. IEEE Communications Letters, 2018, 22(8): 1576-1579.

[199] LI Jinming, PENG Laixian, YE Yilei, et al. A Neighbor Discovery Algorithm in Network of Radar and Communication Integrated System[C]// Proceedings of IEEE 17th International Conference on Computational Science and Engineering (CSE), Chengdu, China: IEEE Press, 2014: 1142-1149.

[200] LIU Ningxin, PENG Laixian, XU Renhui, et al. Neighbor Discovery in Wireless Network with Double-Face Phased Array Radar[C]// Proceedings of 12th International Conference on Mobile Ad-Hoc and Sensor Networks (MSN), Hefei, China: IEEE Press, 2016: 434-439.

[201] LIU Zhenyu, ZHANG Lin, and DING Zhi. Exploiting Bi-Directional Channel Reciprocity in Deep Learning for Low Rate Massive MIMO CSI Feedback [J]. Wireless Communications Letters, 2019, 8(3): 889-892.

[202] WANG Yue, XU Ping, and TIAN Zhi. Efficient Channel Estimation for Massive MIMO Systems via Truncated Two-Dimensional Atomic Norm Minimization [C]// Proceedings of 2017 IEEE International Conference on Communications (ICC), Paris, France: IEEE Press, 2017: 1-6.

[203] PRELCIC N G, RIAL R M and ROBERT W. H Jr. Radar Aided Beam Alignment in MmWave V2I Communications Supporting Antenna Diversity [C]// Proceedings of 2016 Information Theory and Applications Workshop (ITA), La Jolla, CA, USA: IEEE Press, 2016: 1-7.

[204] WAN Xiang, CHEN Tianyi, CHEN Xiaoqing, et al. Beam Forming of Leaky Waves at Fixed Frequency Using Binary Programmable Metasurface[J]. Transactions on Antennas and Propagation, 2018, 66(9): 4942-4947.

[205] ZHANG Yu, LI Qingyu, HUANG Ling, et al. Optimal Design of Cascade LDPC-CPM System Based on Bionic Swarm Optimization Algorithm[J]. Transactions on Broadcasting, 2018, 64(3): 762-770.